\documentclass[12pt]{article}
\usepackage{amsmath}     
\usepackage{citesort}     
\usepackage{amssymb}    
\usepackage{epsfig}  
\def \BE {\begin{equation}}
\def \EE {\end{equation}}
\def \BEA {\begin{eqnarray}}
\def \EEA {\end{eqnarray}}

\def \e {{\epsilon}}

\begin{document}

\title{Discreteness and its effect on the water-wave turbulence.}
\author{Yuri V. Lvov$^1$, Sergey Nazarenko$^2$, Boris Pokorni$^1$
\\ \ \\
{\small $^1$ 
Department of Mathematical Sciences, Rensselaer Polytechnic Institute,Troy NY 12180}\\ 
{\small $^2$ Mathematics Institute, The University of Warwick, Coventry CV4 7AL, UK}}
\maketitle

\begin{abstract}
We perform numerical simulations of the dynamical equations for free
water surface in finite basin in presence of gravity.  Wave Turbulence
(WT) is a theory derived for describing statistics of weakly nonlinear
waves in the infinite basin limit.  Its formal applicability condition
on the minimal size of the computational basin is impossible to
satisfy in present numerical simulations, and the number of wave
resonances is significantly depleted due to the wavenumber
discreteness.  The goal of this paper will be to examine which WT
predictions survive in such discrete systems with depleted resonances
and which properties arise specifically due to the discreteness
effects.  As in \cite{DKZ,onorato,naoto}, our results for the wave
spectrum agree with the Zakharov-Filonenko spectrum predicted within
WT. We also go beyond finding the spectra and compute probability
density function (PDF) of the wave amplitudes and observe an
anomalously large, with respect to Gaussian, probability of strong
waves which is consistent with recent theory \cite{clnp,cln}.  Using a
simple model for quasi-resonances we predict an effect arising purely
due to discreteness: existence of a threshold wave intensity above
which turbulent cascade develops and proceeds to arbitrarily small
scales.  Numerically, we observe that the energy cascade is very
``bursty'' in time and is somewhat similar to sporadic sandpile
avalanches. We explain this as a cycle: a cascade arrest due to
discreteness leads to accumulation of energy near the forcing scale
which, in turn, leads to widening of the nonlinear resonance and,
therefore, triggering of the cascade draining the turbulence levels
and returning the system to the beginning of the cycle.
\end{abstract}

\section{Introduction}

Normal state of the sea surface is chaotic with a lot of waves at
different scales propagating in random directions. Such a state is
referred to as Wave Turbulence (WT). Theory of WT was developed by
finding a statistical closure based on the small nonlinearity and on
the Wick splitting of the Fourier moments, the later procedure is
often interpreted as closeness of statistics to Gaussian or/and to
phase randomness (the two are not the same, see \cite{ln,clnp,cln}).
This closure yields a wave-kinetic equation (WKE) for the waveaction
spectrum.  Such WKE for the surface waves was first derived by
Hasselmann \cite{hasselman}.  A significant achievement in WT theory
was to realize that the most relevant states in WT are energy cascades
through scales similar to the Kolmogorov cascades in Navier-Stokes
turbulence, rather than thermodynamic equilibria as in the statistical
theory of gases.  This understanding came when Zakharov and Filonenko
found an exact power-law solution to WKE which is similar to the
famous Kolmogorov spectrum \cite{Zakfil}.

Numerical simulation of the moving water surface is a challenging
problem due to a tremendous amount of computing power required for
computing weakly nonlinear dispersive waves. This arises due to
presence of widely separated spatial and time scales. As will be
explained below, the weaker we take nonlinearity, the larger we should
take the computational box in order to overcome the $k$-space
discreteness and ignite wave resonances leading to the energy cascade
through scales. In order to maximize the inertial range, one tries to
force at the lowest wavenumbers possible, but without forgetting that
the forcing should be strong enough for the resonance broadening to
overcome the discreteness effect.  But nonlinearity tends to grow
along the energy cascade toward high $k$'s \cite{rough,biven} and,
therefore, the forcing at low wavenumbers should not be too strong for
the nonlinearity to remain weak throughout the inertial range. A
simple estimate \cite{sandpile} says that for the resonant
interactions to be fully efficient, one must have a $N \times N$
computational box with the number of modes $N$ related to the mean
surface angle $\alpha$ as $$N > 1/\alpha^4.$$ Thus, for small enough
nonlinearity $\alpha \sim 0.1$ one must have at least $10000 \times
10000$ resolution which is far beyond present computational capacity.
Thus, in none of the existing numerical experiments, nor in near
future, the formal applicability conditions of WT can be realized. On
the other hand, a small fraction of the resonances can be activated at
levels of nonlinearity which is much less than in the above estimate
and this reduced set of resonant modes can, in principle, be
sufficient to carry the turbulent cascade through scales.  In this
paper we estimate the minimal resonance broadening which is sufficient
to generate the cascade.

It is necessary to examine which of the WT predictions survive beyond
the formal applicability conditions when the cascade is carried by a
depleted set of resonant modes, and which specific features arise due
to such resonance depletion.

As in other recent numerical experiments~\cite{DKZ,onorato,naoto},
here we observe formation of a spectrum consistent with the ZF
spectrum corresponding to the direct energy cascade.  For this, the
resonance broadening at the forcing scale should be maintained at
about an order of magnitude larger than the minimal level necessary
for triggering the cascade.  Further, in agreement with more recent WT
predictions about the higher-order statistics \cite{clnp,cln}, we
observe an anomalously high (with respect to Gaussian) probability of
the large-amplitude waves.  Also in agreement with recent WT findings
\cite{cln}, we observe a buildup of strong correlations of the wave
phases $\phi_k$ whereas the factors $e^{i\phi_k}$ remain
de-correlated.

There are also distinct features arising due to discreteness.  We
analyze them by exploiting the two-peak structure of the time-Fourier
transform at each $k$: a dominant peak at (very close to) linear
frequency $\omega_k$, and a weaker one with a frequency approximately
equal to $2 \omega_{k/2}$. The second peak is a contribution of the
$k/2$-mode in the nonlinear term of the canonical transformation
relating the normal variable and the observables (e.g. surface
elevation). In fact, the nature of the second frequency peak is quite
well understood in literature and it has even been used for remote
sensing of vertical sheer by VHF frequency radars \cite{shrira}.  In
absence of nonlinearity one would observe only the first but not the
second peak and, therefore, one can quantify the nonlinearity level as
the ratio of the amplitudes of these two peaks.  When the second peak
becomes stronger than the first one, the wave phase experiences a
rapid and persistent monotonic change.  Detecting such phase ``runs''
gives an interesting picture of the nonlinear activity in the 2D
$k$-space.  In particular, we notice a ``bursty'' nature of the energy
cascade resembling sandpile avalanches. Possible explanation of such
behavior is the following. When nonlinearity is weak, there is no
wave-wave resonances, consequently there is no effective energy
transfer, and system behaves like ``frozen turbulence'' (term
introduced in \cite{frozen} for the capillary wave turbulence).
Energy generated at the forcing scale will accumulate near this scale
and the nonlinearity will grow.  When resonance broadening gets wide
enough, so that the resonances are not inhibited by discreteness, the
nonlinear wave-wave energy transfer starts, which diminishes
nonlinearity and subsequently ``arrests'' resonances. Thus the system
oscillates between having almost linear oscillations with stagnated
energy and occasional avalanche-like discharges.

%
\section{Equations for the free surface}
%

Let us consider motion of a water volume of infinite depth embedded in
gravity and bounded by a surface separating it from air at height $z =
\eta({\bf x}, t)$ where ${\bf x} = (x,y)$ is the horizontal
coordinate. Let the velocity field be irrotational, ${\bf u} = \nabla
\Phi$, so that the incompressibility condition becomes
\begin{equation}
\Delta \Phi = 0, \hspace{1cm} \hbox{for\;\;} z < \eta({\bf x}, t).
\label{laplase}
\end{equation}
Rate of change of the surface elevation must be equal to the vertical
velocity of the fluid particle on this surface, which gives
\begin{equation}
D_t \eta = \partial_z \Phi,
\hspace{1cm} \hspace{1cm} \hbox{for\;\;} z = \eta({\bf x}, t),
\label{zveloc}
\end{equation}
where $D_t = \partial_t + {\bf u} \cdot \nabla_\perp $ is the material
time derivative.  The second condition at the surface arises from the
Bernoulli equation in which pressure is taken equal to its atmospheric
value. This condition gives
\begin{equation}
\partial_t \Phi + {1 \over 2} | \nabla \Phi|^2 = - g \eta,
\hspace{1cm} \hspace{1cm} \hbox{for\;\;} z = \eta({\bf x}, t),
\label{bernoul}
\end{equation}
where $g$ is the free-fall acceleration.

Although equations (\ref{zveloc}) and (\ref{bernoul}) involve only
two-dimensional coordinate ${\bf x}$, the system remains
three-dimensional due to the 3D equation (\ref{laplase}). One can
transform these equations to a truly 2D form by assuming that the
surface deviates from its rest plane only by small angles and by
truncating the nonlinearity at the cubic order with respect to the
small deviations.  This procedure yields the following dynamical
equations (see e.g.~\cite{zakheqn,Choi}):
\begin{eqnarray}
\eta _{t} &=&\mathbf{\Gamma }\left[ \Psi \right]  \notag \\
&&-\varepsilon \left( \mathbf{\Gamma }\left[ \mathbf{\Gamma }\left[ \Psi %
\right] \eta \right] +\nabla _{\bot }\cdot \left[ \left( \nabla _{\bot }\Psi
\right) \eta \right] \right) \label{DEeta} \\
&&+\varepsilon ^{2}\left( \Gamma \left[ \mathbf{\Gamma }\left[ \mathbf{%
\Gamma }\left[ \Psi \right] \eta \right] \eta \right] +\frac{1}{2}\Gamma %
\left[ \left( \Delta _{\bot }\Psi \right) \eta ^{2}\right] +\frac{1}{2}%
\Delta _{\bot }\left( \mathbf{\Gamma }\left[ \Psi \right] \eta ^{2}\right)
\right), \notag \\
\Psi _{t} &=&-g\eta  \notag \\
&&-\varepsilon \frac{1}{2}\left( \left| \nabla _{\bot }\Psi \right|
^{2}-\left( \Gamma \lbrack \Psi ]\right) ^{2}\right) \label{DEphi} \\
&&-\varepsilon ^{2}\Gamma \lbrack \Psi ]\left( \mathbf{\Gamma }\left[ 
\mathbf{\Gamma }\left[ \Psi \right] \eta \right] +\left( \Delta _{\bot }\Psi
\right) \eta \right),  \notag
\end{eqnarray}
where
\begin{equation}
 \Psi = \Phi |_{z = \eta({\bf x}, t)},
\end{equation}
and $\Gamma$ is the Gilbert transform which in the Fourier space
corresponds to multiplication by $ k = |{\bf k}|$, i.e.
\begin{equation*}
\Gamma \left[ f\right] (\mathbf{x},t)=\frac{1}{2\pi }\int k
\widehat{f}(\mathbf{k}%
,t)e^{i\mathbf{k}\cdot \mathbf{x}}d\mathbf{k}.
\end{equation*}
Here, we have the following convention for defining the Fourier
transform
\BE
\widehat{f}(\mathbf{k}) = {1 \over 2 \pi} \int e^{i (\mathbf{k} \cdot 
\mathbf{x})} f(\mathbf{x}) \, d \mathbf{k}.
\EE
In equations (\ref{DEeta}) and  (\ref{DEphi}), we rescaled variables
$\eta $ and $\Psi$ to make them order one, so that the nonlinearity 
smallness is now in a formal parameter $\epsilon \ll 1$.

Truncated equations (\ref{DEeta}) and (\ref{DEphi}) will be used for
our numerical simulations. They have a convenient form for the
pseudo-spectral method which computes evolution of the Fourier modes
but switches back to the coordinate space for computing the nonlinear
terms.  However, for theoretical analysis these equations have to be
diagonalised in the $k$-space and a near-identity canonical
transformation must be applied to remove the nonlinear terms of order
$\epsilon$ since the gravity wave dispersion $\omega = \sqrt{gk}$ does
not allow three-wave resonances. The resulting equation is also
truncated at $\epsilon^2$ order and it is called the Zakharov equation
\cite{zakheqn,Zak91,ZakAMS98,ZLF},
\begin{equation}
i \dot a_l = \epsilon^2 \sum_{\alpha\mu\nu}
{ W^{l\alpha}_{\mu\nu}}\bar a_\alpha a_\mu a_\nu 
e^{i\omega^{l\alpha}_{\mu\nu}t}
\delta^{l\alpha}_{\mu\nu},
\label{FourWaveEquationOfMotionB}
\end{equation}
where $\omega^{l\alpha}_{\mu\nu} =
\omega_l+\omega_{\alpha}-\omega_{\mu}-\omega_{\nu}$,
$\omega_l = \sqrt{g k_l}$ is the frequency of mode $ {\bf l}$
($\; {\bf l} \in {\cal Z}^2$), ${\bf k}_l = 2\pi {\bf l} /L$ is the wavenumber,
$L$ is the box size and $k_l = |{\bf k}_l|$.
Here, $a_l$ is the wave action variable in the interaction
 representation, $a_l = e^{i \omega_l t} b_l$ where  $b_l$ 
is a normal variable,
\begin{equation}
b_l = \sqrt{\omega_l \over 2 k_l} \eta_l + i \sqrt{k_l \over 2 
\omega_l} \Psi_l +0(\epsilon).
\label{nv}
\end{equation}
Here, $0(\epsilon)$ terms appear because of the near-identity
canonical transformation needed to remove the quadratic terms from the
evolution equations \cite{zakheqn,krasitskii,ZLF,PRZ,Zak91,ZakAMS98}.
Expression for the interaction coefficient $W^{l \alpha}
_{\mu \nu}$ is lengthy and can be found in \cite{krasitskii}.

Zakharov equation is of fundamental importance for theory and it is
also sometimes used for numerics.  However, in our work we choose to
compute equations (\ref{DEeta}) and (\ref{DEphi}) because this allows
us to use the standard trick of pseudo-spectral methods via computing
the nonlinear term in the real thereby accelerating the code.

\section{Statistical Quantities in Wave Turbulence }

Let us consider a wavefield in a periodic square basin of side $L$ and
let the Fourier representation of this field be $a_l(t)$ where index
${\bf l} {\in } {\cal Z}^2$ marks the mode with wavenumber ${\bf k}_l
= 2 \pi {\bf l} /L$ on the grid in the $2$-dimensional Fourier space.
Discrete $k$-space is important for formulating the statistical
problem.  For simplicity let us assume that there is a cut-off
wavenumber $k_{max}$ so that thee is no modes with wavenumber
components greater than $k_{max}$, which is always the case in
numerical simulation.  In this case, the total number of modes is $N =
(k_{max} / \pi L)^2$ and index ${\bf l}$ will only take values in a finite
box, ${\bf l} \in {\cal B}_N \subset {\cal Z}^2$ which is centered at 0 and
all sides of which are equal to $N^{1/2}$.  To consider homogeneous
turbulence, the large box (i.e. continuous $k$) limit, $N \to \infty
$, will have to be taken later.

Let us write the complex $a_l$ as $a_l =A_l \psi_l $ where $A_l$ is a
real positive amplitude and $\psi_l $ is a phase factor which takes
values on ${\cal S}^{1} $, a unit circle centered at zero in the
complex plane.  The most general statistical object in WT \cite{cln}
is the $N$-mode joint PDF ${\cal P}^{(N)}$ defined as the probability
for the wave intensities $A_l^2 $ to be in the range $(s_l, s_l +d
s_l)$ and for the phase factors $\psi_l$ to be on the unit-circle
segment between $\xi_l$ and $\xi_l + d\xi_l$ for all ${\bf l} \in
{\cal B}_N$.

The fundamental statistical property of the wavefield in WT is that
all the amplitudes $A_l$ and phase factors $\psi_l $ are independent
statistical variables and that  all $\psi_l $'s are uniformly 
distributed on ${\cal S}^{1} $. 
This kind of statistics was introduced in \cite{ln,clnp,cln} and called ``Random
Phase and Amplitude'' (RPA) field.
In terms of the PDF, we say that the field $a$ is of
RPA type if it can be product-factorized,
\BE
{\cal P}^{(N)} \{s, \xi \}  = {1 \over (2 \pi)^{N} } \prod_{ {\bf l}
\in  {\cal  B}_N } 
 P^{(a)}_{l} (s_l),
\EE
where $ P^{(a)}_{l}(s_l)$ is the one-mode PDF for variable $A_l^2$.

Note that in this formulation 
the distributions of
$A_l$ remain unspecified and, therefore, the amplitudes do not have to be deterministic
(as in earlier works using RPA) nor do they have to correspond to Gaussianity,
 \BE
 P^{(a)}_{l}(s_l)  = {1 \over n_l} \, \exp{(-s_l/n_l)}
\EE
where $ n_l = \langle s_l \rangle $ is the waveaction spectrum.

Importantly, RPA formulation involves independent {\em phase factors}
$\psi = e^{i \phi}$ and not {\em phases} $\phi$. Firstly, the phases
would not be convenient because the mean value of the phases is
evolving with the rate equal to the nonlinear frequency correction
\cite{cln}. Thus one could not say that they are ``distributed
uniformly from $-\pi$ to $\pi$''. Moreover the mean fluctuation of the
phase distribution is also growing and they quickly spread beyond
their initial $2 \pi$-wide interval \cite{cln}.  But perhaps even more
important, it was shown in \cite{cln} that $\phi$'s build mutual
correlations on the nonlinear timescale whereas $\psi$'s remain
independent.  In the present paper we are going to check this
theoretical prediction numerically by directly measuring the
properties of $\phi$'s and $\psi$'s.

In \cite{ln,clnp} RPA was {\em assumed} to hold over the nonlinear
time.  In \cite{cln} this assumption was examined {\em a posteriori},
i.e. based on the evolution equation for the multi-point PDF. Note
that only the phase randomness is necessary for deriving this
equation, whereas both the phase and the amplitude randomness are
required for the WT closure for the one-point PDF or the kinetic
equation for the spectrum.  This fact allows to prove that, if valid
initially, the RPA properties survive in the leading order in small
nonlinearity and in the large-box limit \cite{cln}. Such an
approximate leading-order RPA is sufficient for the WT closure.

\section{Theoretical WT predictions}

When the wave amplitudes are small, the nonlinearity is weak and the
wave periods, determined by the linear dynamics, are much smaller than
the characteristic time at which different wave modes exchange
energy. In the other words, weak nonlinearity results in a timescale
separation and this fact is exploited in WT to describe the slowly
changing wave statistics by averaging over the fast linear
oscillations.


\subsection{Evolution equations for the PDF's, moments and spectrum.}


In \cite{cln}  the following
equation for 
 for the $N$-mode PDF was obtained for  the four-wave systems,
\def \k {{\bf k}}
\BE
\dot  {\cal P} = { \pi \e^2 } \int |W^{jl}_{nm}|^2\delta(\tilde\omega^{jl}_{nm})
\delta^{jl}_{nm}
\left[{\delta  \over \delta s} \right]_4 \left(s_j s_ls_m s_n 
\left[{\delta \over \delta s} \right]_4 {\cal P} \right)  \, d\k_j  d\k_l d\k_m d\k_n,
\label{peierls4}
\EE
where 
\BE
\left[{\delta  \over \delta s} \right]_4 = {\delta  \over \delta s_j}+{\delta  \over \delta s_l} - {\delta
  \over \delta s_m} -{\delta  \over \delta s_n}.
\label{brac4}
\EE 
Here $N \to \infty$ limit has already been taken and ${\delta
\over \delta s_j}$ means the variational derivative.  Using this
equation, one can prove that RPA property holds over the nonlinear
time, i.e. the $N$-mode PDF remains of the product factorized form
with accuracy sufficient for the WT closures to work \cite{cln}.
Using RPA, we get for the one-point marginals \cite{cln},
\begin{equation}
{\partial P_a \over \partial t}+  {\partial F \over \partial s_j}  =0,
 \label{pa}
\end{equation}
 with $F$ is a probability flux in the s-space,
\begin{equation}
F=-s_j (\gamma P_a +\eta_j {\delta P_a \over \delta s_j}),
\label{flux1}
\end{equation}
where 
\begin{eqnarray}
\eta_j &=& 4 \pi \epsilon^2 \int
|W^{jl}_{nm}|^2 \delta^{jl}_{nm}
\delta(\omega^{jl}_{nm}) n_l n_m n_n \,  d { \k_l} d { \k_m}  d { \k_n,} 
\label{RHO1} \\
\gamma_j &=&
4 \pi \epsilon^2 \int  |W^{jl}_{nm}|^2 \delta^{jl}_{nm}
\delta(\omega^{jl}_{nm})
 \Big[ n_l (n_m + n_n) - n_m n_n\Big] \,  d { \k_l} d { \k_m}  d { \k_n.}
 \label{GAMMA1}
\end{eqnarray}
Here we introduced the wave-action spectrum,
\BE
n_j = \langle A_j^2 \rangle.
\EE 
From (\ref{pa}) we get the following equation
for the  moments
$ M^{(p)}_j = \langle A_j^{2p} \rangle $:
\BE
\dot M^{(p)}_j = -p \gamma_j M^{(p)}_j +
p^2 \eta_j M^{(p-1)}_j.\label{MainResultOne} 
\EE
which, for $p=1$ gives the standard wave kinetic equation (WKE),
\BE
 \dot n_j = - \gamma_j n_j + \eta_j . 
\label{ke}
\EE


\subsection{Preservation of the RPA property.}


Validity of the WT theory relies on persistence of the RPA property of
the wavefield over the nonlinear evolution time.  Such persistence was
demonstrated in \cite{cln} based on the evolution equation for the
multi-mode PDF, where the product factorization of PDF was shown to
hold with an accuracy sufficient for the WT closure.  It was also
emphasized in \cite{cln} that RPA must use independent phase factors
$\psi_k$ rather than the phases $\phi_k$ independence of which does
not survive over the nonlinear time.  The theoretical prediction of
persistent independence of $A_k$'s and $\psi_k$'s and about the growth
of correlations of $\phi_k$'s will be checked in this paper
numerically.

Further, the WT approach predicts that the mean value of the phase
grows with a rate given by the nonlinear frequency correction and that
the r.m.s. fluctuation of the phase also grows in time \cite{cln}.  In
this paper we will see that in reality the time evolution of the phase
is more complicated than this WT prediction: the phase exhibits
quasi-periodic fluctuations intermittent by rare ``phase runs'', -
monotonic changes over several linear periods by large values which
can significantly exceed $2 \pi$.
 

\subsection{Steady state solutions.}


Steady power-law solutions of WKE which correspond to 
a direct cascade of energy  and an inverse waveaction cascade are,
\begin{eqnarray}
n(k) &=&C_{1}P^{1/3}k^{-4} \label{direct} \\ n(k)
&=&C_{2}Q^{1/3}k^{-23/6} \label{inverse},
\end{eqnarray}
where $P$ and $Q$ are the energy and the waveaction fluxes
respectively and $C_1$ and $C_2$ are constants, and $k=|{\bf k}|$.
The first of these solutions is the famous ZF spectrum \cite{Zakfil}
and it has a great relevance to the small-scale part of the sea
surface turbulence.  It has been confirmed in a number of recent
numerical works \cite{DKZ,onorato,naoto}, but we will also confirm it
in our simulation.

Now, let us consider the steady state solutions for the one-mode PDF.
Note that in the steady state $\gamma /\eta = n$ which follows from
WKE (\ref{ke}).  Then, the general steady state solution to (\ref{pa})
is \BE P=\hbox{const} \, \exp{(-s/n)} -({F}/{\eta}) Ei({s}/{n})
\exp{(-s/n)}, \EE where $Ei(x)$ is the integral exponential function.
At the tail $s \gg n_k$ we have
\begin{equation}
P \to - \frac{F}{s\gamma}
\label{Ppart}
\end{equation}
if $F \ne 0$.  The $1/s$ tail decays much slower than the exponential
(Rayleigh) part and, therefore, it describes strong intermittency.  On
the other hand, $1/s$ tail cannot be infinitely long because otherwise
the PDF would not be normalizable.  As it was argued in \cite{cln},
the $1/s$ tail should with a cutoff because the WT description breaks
down at large amplitudes $s$.  This cutoff can be viewed as a
wavebreaking process which does not allow wave amplitudes to exceed
their critical value, $P(s) =0$ for $s > s_{nl}$.

Relation between intermittency and a finite flux in the amplitude
space was observed numerically also for the Majda-Mc-Laughlin-Tabak
model by Rumpf and Biven \cite{benno}.

\section{Resonant interaction in discrete $k$-space}

Importance of the $k$-space discreteness for weakly nonlinear waves in
finite basins were realized by Kartashova \cite{Kartashova} and it was
later discussed in a number of papers, \cite{frozen,CNP,ty,meso}.
Nonlinear wave interactions crucially depend on the sort of resonances
the dispersion relation allows. The dispersion relation of the surface
gravity waves is concave and hence it forbids three-wave interactions,
so that the dominant process is four-wave. Resonant manifold is
defined by resonant conditions
$\mathbf{k}+\mathbf{k}_{1}=\mathbf{k}_{2}+\mathbf{k}_{3}$, $\omega
_{\mathbf{k}}+\omega _{\mathbf{k}_{1}}=\omega _{\mathbf{k}_{2}}+\omega
_{\mathbf{k}_{3}}$ or, substituting $\omega _{\mathbf{k}}=\sqrt{gk}$,
\begin{eqnarray}
\mathbf{k}+\mathbf{k}_{1} &=& \mathbf{k}_{2}+\mathbf{k}_{3} \notag \\
\sqrt{k}+\sqrt{k_1} &=& \sqrt{k_2}+\sqrt{k_3}
\label{RC}
\end{eqnarray}
The problem of finding exact resonances in discrete $k$-space was
first formulated by Kartashova \cite{Kartashova} who gave a detailed
classification of for some types of waves, e.g. Rossby waves. For the
deep-water gravity waves, Kartashova only considered
\textit{symmetric} solutions, i.e. of the form $k=k_2, k_1=k_3$ (or
$k=k_3, k_1=k_2$).  Interestingly, there appear to be also
\textit{asymmetric} solutions,

Solutions for collinear quartets and the tridents are easy to find by 
rewriting the resonant conditions (\ref{RC}) as polynomial equations
(appropeiately re-arranging and taking squares of the equations) and
using rational parametrisations of their solutions.
This way we get the following family of the
collinear quartets \cite{dlz},
$$\mathbf{k}=(a,0),\;\;  \mathbf{k}_1=(b,0),\;\;
\mathbf{k}_2=(c,0),\;\; \mathbf{k}_3=(d,0) $$
with
$$ a = m^2 (m+n)^2, \;\;
b = n^2 (m+n)^2, \;\;
c= -m^2 n^2, \;\;
d= (m+n)^4 + m^2 n^2,
$$
where $m$ and $n$ are natural numbers, and ``tridents'',
$$\mathbf{k}=(a,0),\;\;  \mathbf{k}_1=(-b,0),\;\;
\mathbf{k}_2=(c,d),\;\; \mathbf{k}_3=(c,-d) $$
with
$$ a = (s^2+t^2+st)^2, \;\;
b = (s^2+t^2-st)^2,  \;\;
c= 2st(s^2+t^2), \;\;
d= s^4-t^4,
$$ where $s$ and $t$ are integers.  Both of these classes can be
easily extended by re-scaling, i.e. multiplying all four vectors by an
integer.  Even more solutions can be obtained via rotation by an angle
with rational-valued cosine (this gives a new rational solution) and
further re-scaling (to obtain integer solution out of the rational
one).

There are other, rather rare, exact nontirvial resonances, for example
one provided to us by Kartashova
\cite{PrivateCommunicationKartashova}: $${\bf k}=(495,90),\ {\bf
k_1}=(64,128), \ {\bf k_2}=(359,118), \ {\bf k_3}= (200,100). $$ The
complete investigation of the exact resonance types and their
respective roles is a facinating subject of future work).

Note that the interactions coefficient vanishes on the collinear
quartets \cite{fivewave} and, therefore, these quartets do not
contribute into the turbulence evolution.  Secondly, there appears to
be much more symmetric quartets than tridents, so the later are
relatively unimportant for the nonlinear dynamics too.

\begin{figure}
\begin{center}
\includegraphics[width=5cm]{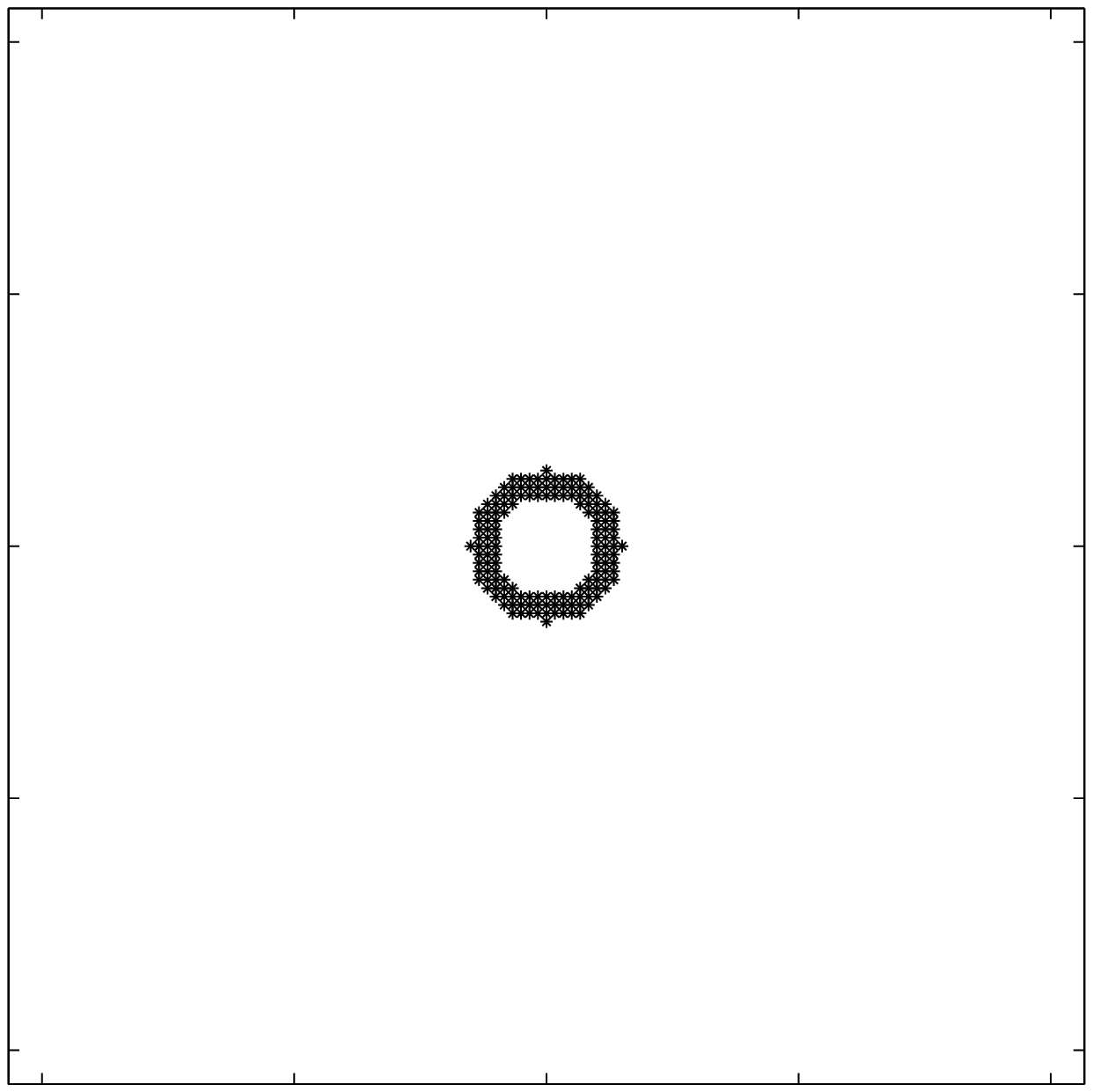}
\includegraphics[width=5cm]{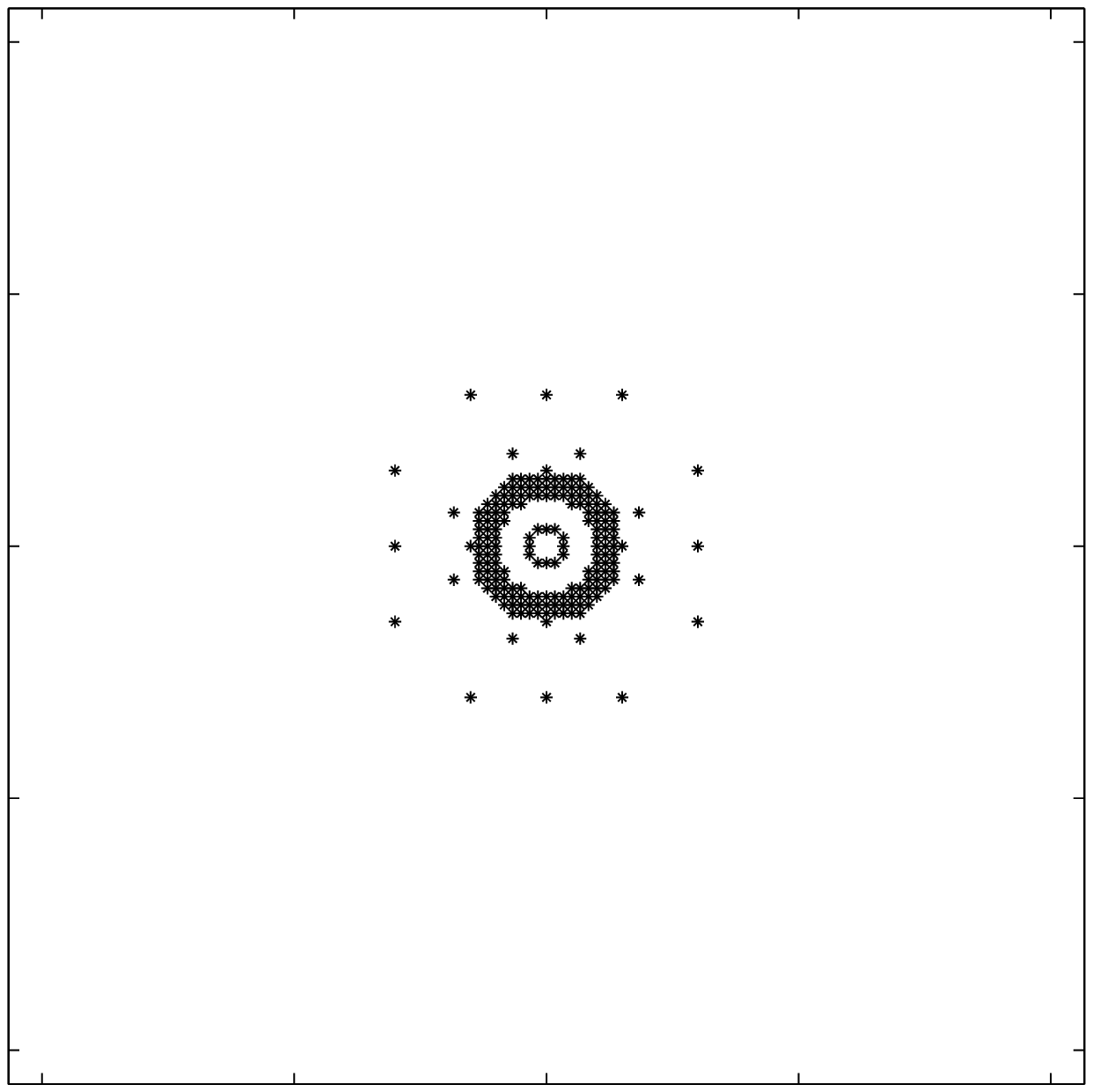}
\includegraphics[width=5cm]{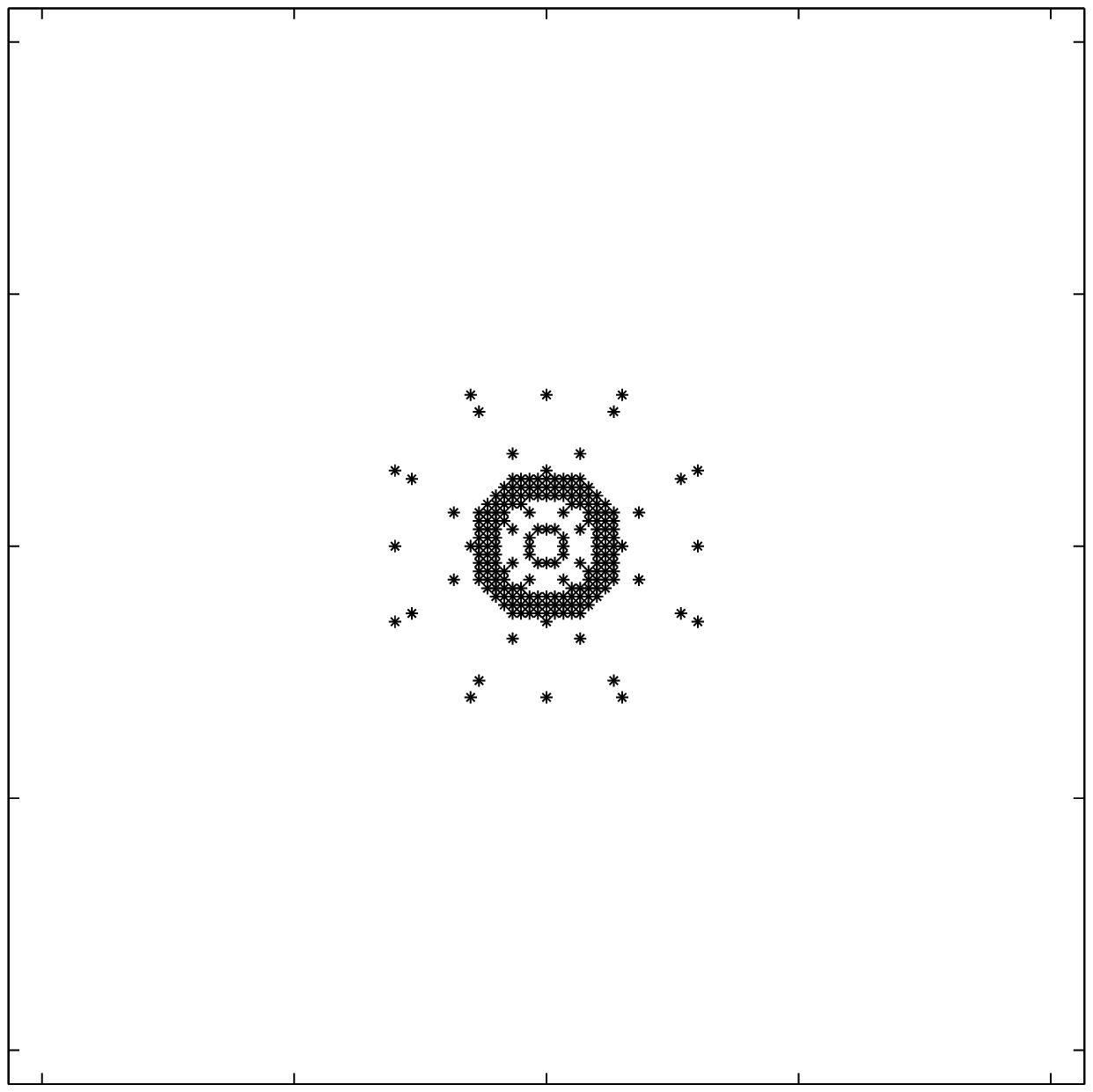}\\
\includegraphics[width=5cm]{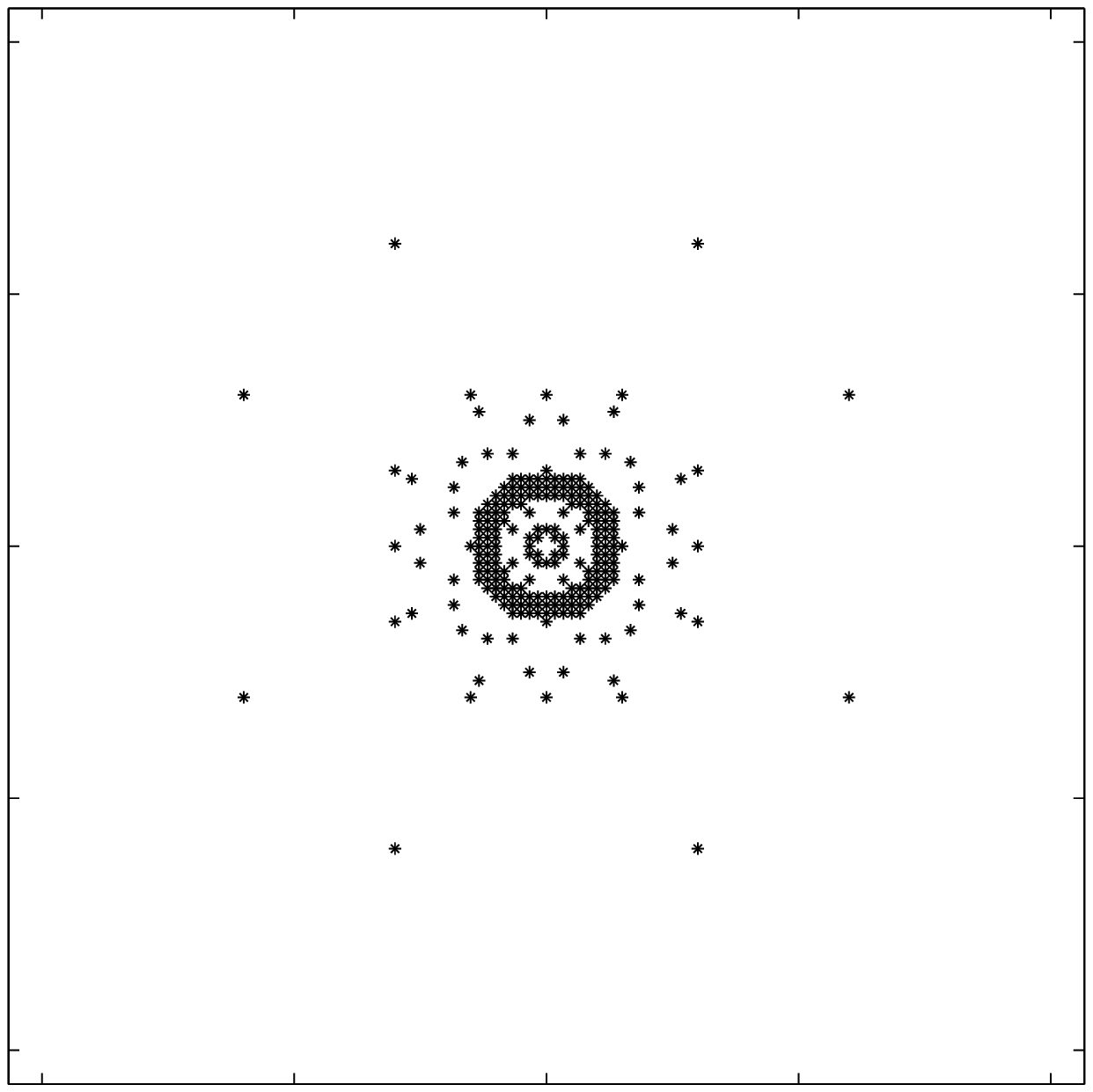}
\includegraphics[width=5cm]{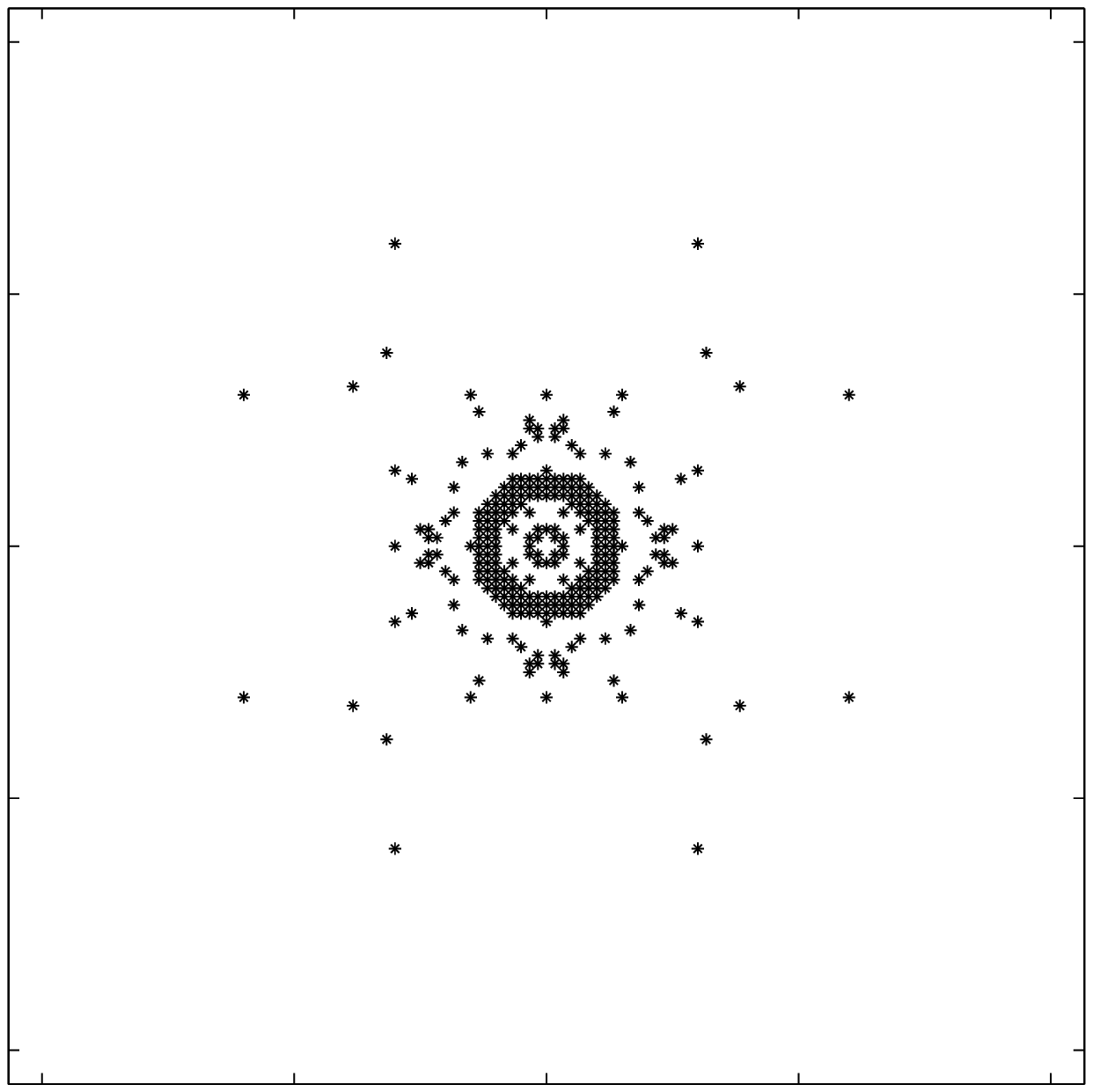}
\includegraphics[width=5cm]{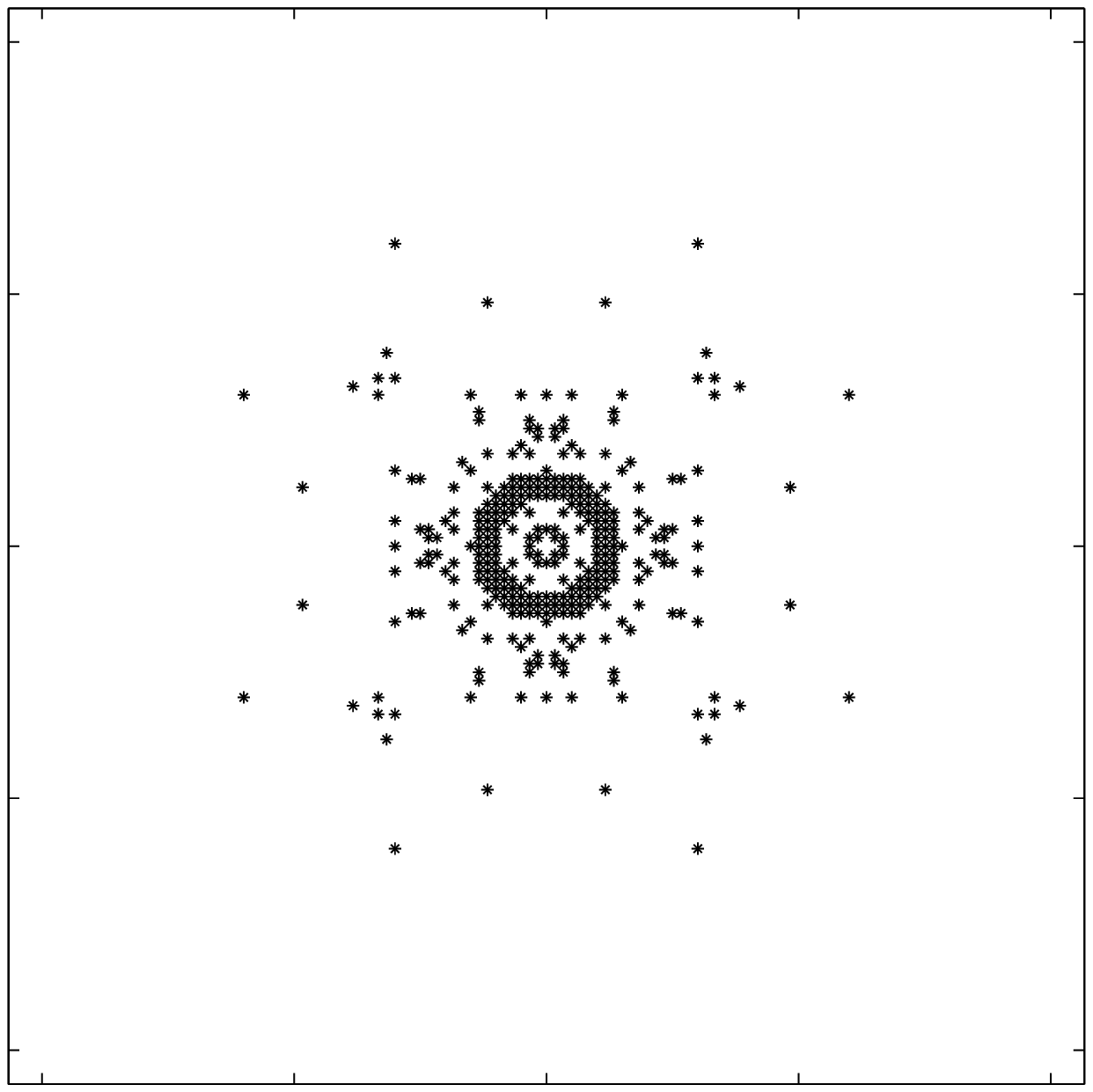}\\
\includegraphics[width=5cm]{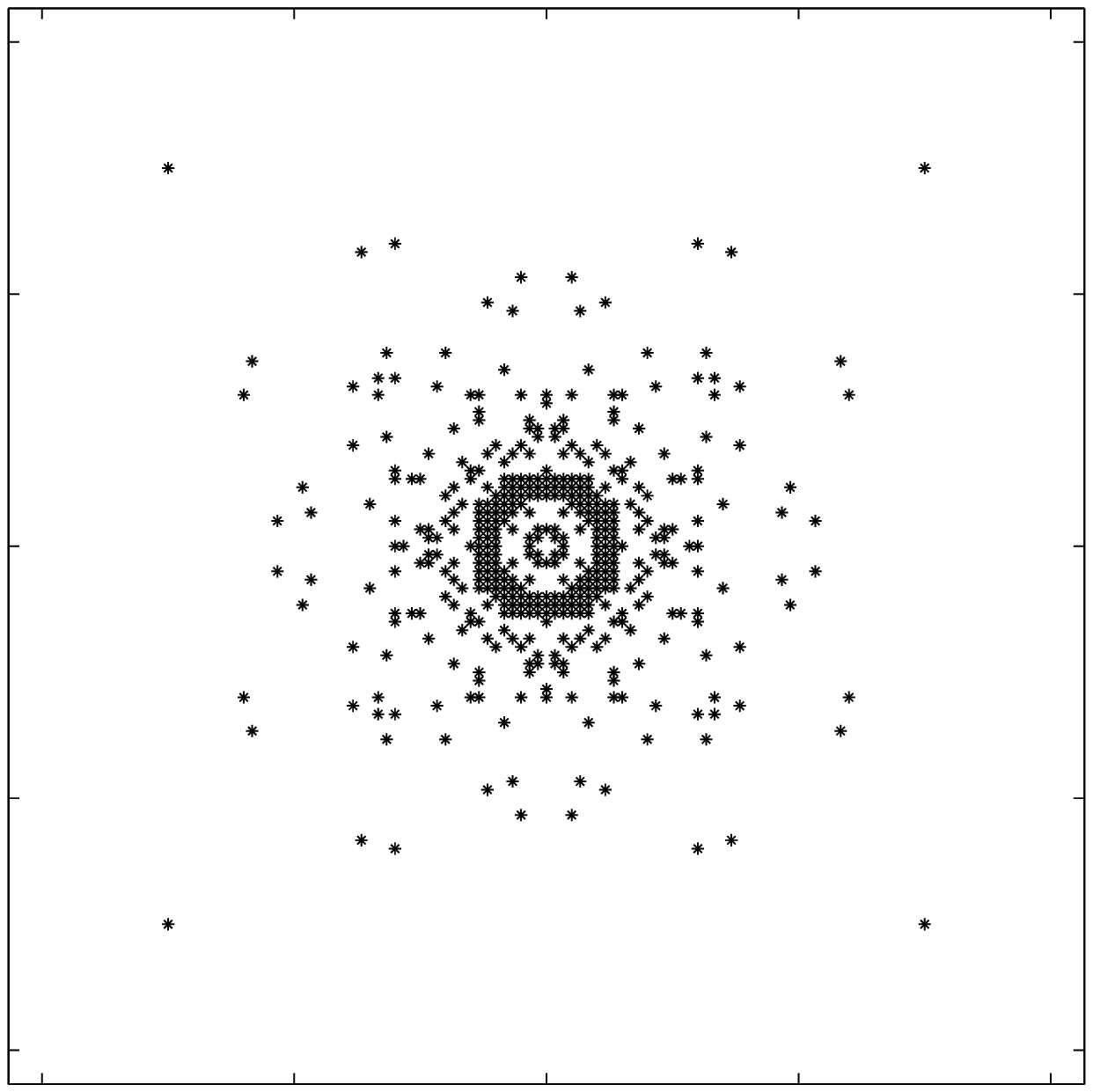}
\includegraphics[width=5cm]{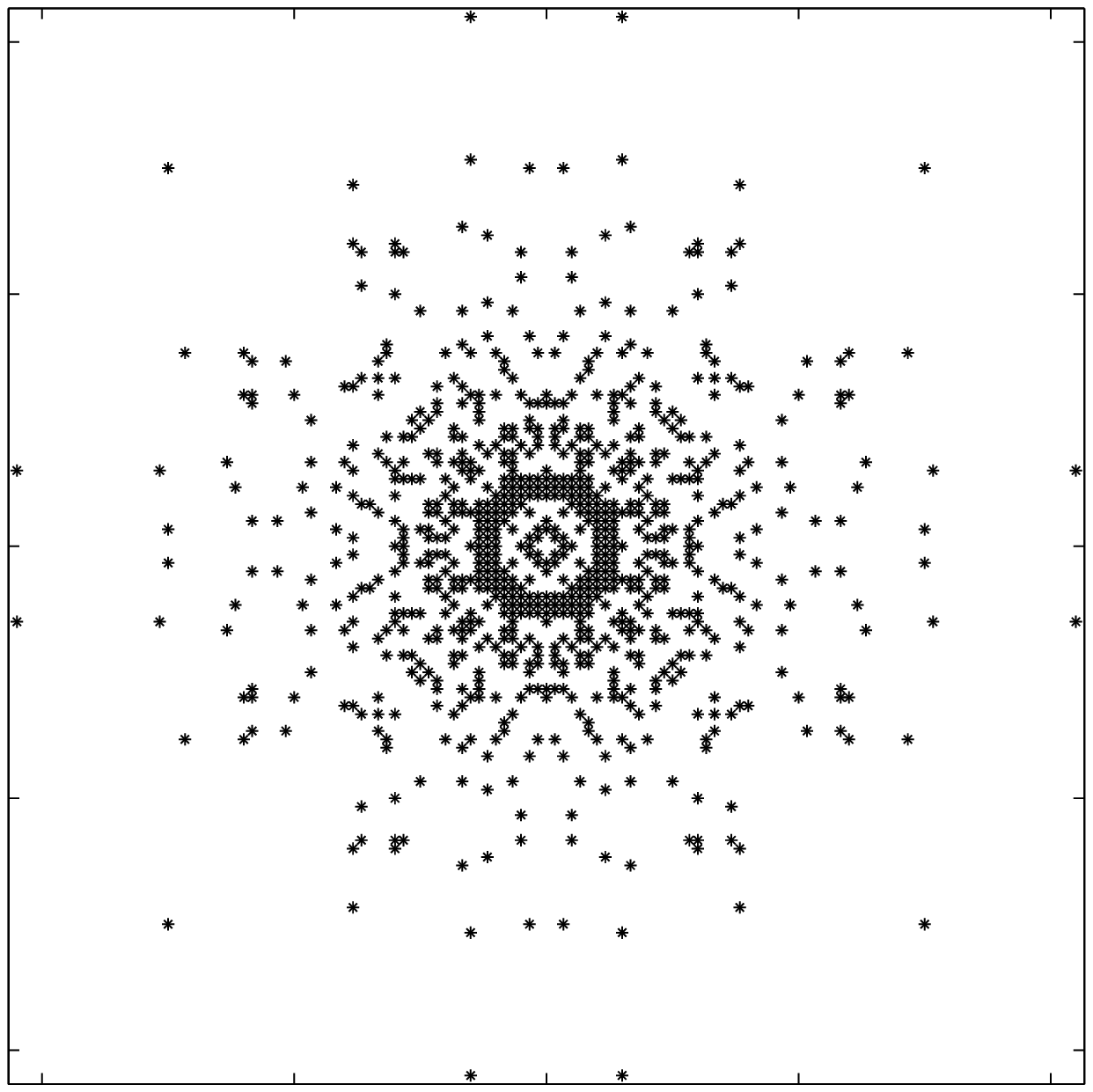}
\includegraphics[width=5cm]{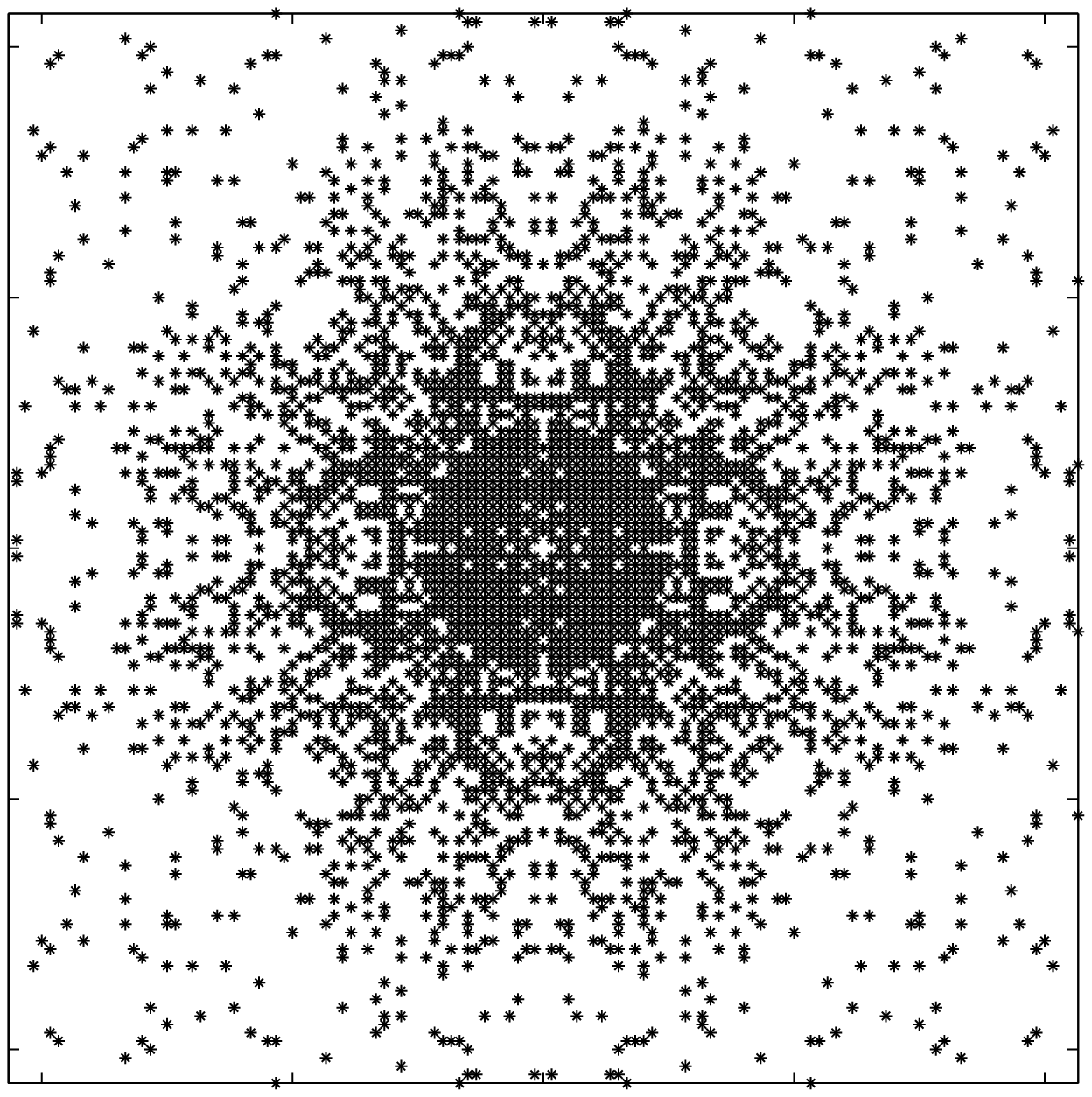}
\caption[6pts]{ Quasi-resonant generations of modes on Fourier space
 $[-64,64]^{2}$, where initially (generation 1) only modes in the ring
 $6<k<9$ were present. 
Each next generation consists of the union of the modes of the current
 generation and the new modes which satisfy the quasi-resonance
condition with the modes of the current generation.
Here the value of the broadening is slightly above
 $\delta_{crit}$}
\label{ResGen}
\end{center}
\end{figure}

Because of nonlinearity, the wave resonances have a finite width.
Even though this width is small in weakly nonlinear systems, it may be
sufficient for activating new mode interactions on the $k$-space grid
and thereby trigger the turbulent cascade through scales.  Such
quasi-resonances can be roughly modeled through
$\mathbf{k}_{1}+\mathbf{k}_{2}=\mathbf{k}_{3}+\mathbf{k}_{4}$, $\left|
\omega _{\mathbf{k}_{1}}+\omega _{\mathbf{k}_{2}}-\omega
_{\mathbf{k}_{3}}-\omega _{\mathbf{k}_{4}}\right| <\delta $, where
$\delta $ describes the resonance broadening.  Figure \ref{ResGen}
shows quasi-resonant generations of modes on space $[-64,64]^{2}$,
where initially (generation 1) only modes in the ring $6<k<9$ were
present (as in our numerical experiment).  With broadening of the
resonant manifold smaller than the critical $\delta _{crit}\sim
1.4\ast 10^{-5}$, a finite number of modes outside the initial region
get excited due to exact resonances (generation 2) but there are no
quasi-resonances to carry energy to outer regions in further
generations.  If broadening is larger than critical, energy cascades
infinitely.  Contrary to the the case of capillary waves where
quasi-resonant cascades die out if broadening is not large enough
\cite{CNP}, in the case presented here quasi-resonant cascades either
do not happen at all (if $\delta <\delta _{\rm crit}$) or they spread
through the wavenumber space infinitely (if $\delta >\delta _{\rm
crit}$), as happens on Figure (\ref{ResGen}).
%
%

\section{Numerical simulation}

Numerical simulations presented in this work were performed on a
single-processor workstation (2.5GHz, 1Gb RAM). We performed a direct
numerical simulation, integrating the dynamical equations of motion
(\ref{DEeta}) and (\ref{DEphi}) using pseudo-spectral method with
resolution of $256\times256$ wavenumbers. Numerical integrator used
for advancing in time was RK7(8) presented in \cite{RK8}. Time step
was $\frac{T_{min}}{35}$ where $T_{min}$ is the period of the shortest
wave on the axis.  Approximate processor time for this work was 4.5
weeks.

In our numerical experiment, we force the system in the $k$-space ring
$k_* < k < k^*$ with $k_* =6$ and $k^* = 9$.  This ring is located at
the low wavenumber part of the $k$-space in order to generate energy
cascade toward large $k$'s, but we deliberately avoid forcing even
longer waves ($k<6$) because our experience shows that this would lead
to undesirable strong anisotropic effects. In the ring, we fix the
shape to coincide with the ZF spectrum, $<|a_{\mathbf{k}}(t)|^{2}>\sim
k^{-4}$, and hence we set $|\eta _{\mathbf{k} }| \sim k^{-7/4}$,
$|\Psi _{\mathbf{k}}| \sim k^{-9/4}$. These fixed amplitudes were then
multiplied by random phase factors.  Thus, surface $\eta
_{\mathbf{k}}$ and velocity potential $\Psi_{\mathbf{k}}$ were set to
$2\pi ^{3} \, e^{i \theta _{\mathbf{k}}} \ast k^{\alpha
  _{\mathbf{k}}}$,
 where $\theta _{\mathbf{k}}$ were uniformly distributed in $[0,2\pi
 ]$ and
\begin{equation*}
\alpha _{k}=x\left\{ 
\begin{array}{cc}
\left[ 1+\left( \frac{k_{\ast }-k}{k^{\ast }}\right) ^{2}\right] ^{3/4} & 
\hbox{if} \;k\in \left( 0,k_{\ast }\right) \\ 
1 & 
\hbox{if}
\;k\in \lbrack k_{\ast },k^{\ast }] \\ 
\left( \frac{k}{k^{\ast }}\right) ^{3/4} & \hbox{if}
\;k\in \left( k^{\ast },\frac{N%
}{2}\right)
\end{array}
\right\}
\end{equation*}
where $x=-7/4$ for the surface and $x=-9/4$ for velocity potential.
Damping was applied in both small and large wavenumber regions.  At
small wavenumbers inside the forcing ring, we applied an adaptive
damping to prevent formation of undesirable ``condensate'' which could
spoil isotropy and locality of scale interactions.  At large
wavenumbers, damping is needed to absorb the energy cascade and,
therefore, to avoid ``bottleneck'' spectrum accumulation near the
cutoff wavenumber. In our simulations, we implemented the damping as a
low-pass filter $\gamma _{\mathbf{k}}$ applied to the $k$-space
variables at each time step.  The damping function had the form
\begin{equation*}
\gamma _{\mathbf{k}}=\left\{ 
\begin{array}{cc}
5\left( k-6\right) ^{3/2} & k<6 \\ 
0 & k\in \left[ 6,64\right] \\ 
0.028\left( k-64\right) ^{2} & k>64
\end{array}
\right\}
\end{equation*}

 Nonlinearity parameter was set
to $\varepsilon =2\cdot 10^{-2}$, which is a sufficient value to
produce a resonance broadening for supporting energy cascade.

\section{Results}

\subsection{Spectrum}

\begin{figure}
\begin{center}
\includegraphics[width=10cm]{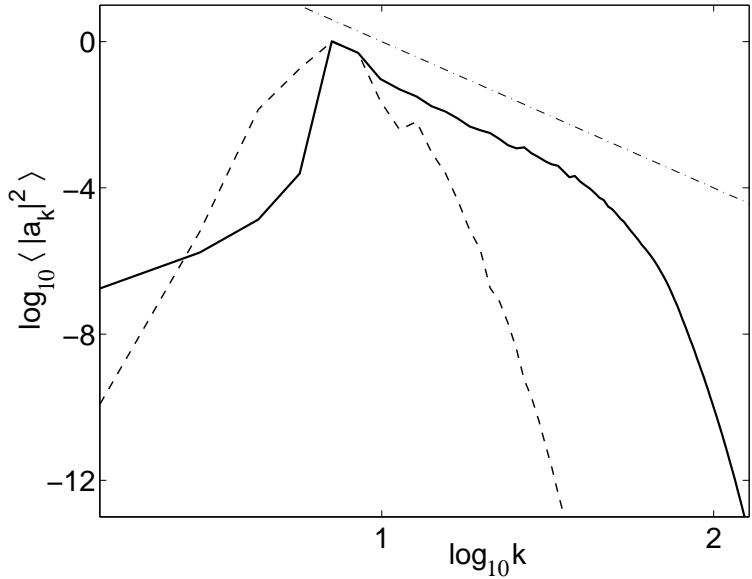}\\
\caption[6pts]{Waveaction spectrum of waves.
Dashed and solid lines show spectra after 300 and 2000
periods of  mode $k=10$.
The straight line corresponds to the ZF spectrum
($-4$ slope). }
\label{EnergySpectrum}
\end{center}
\end{figure}

Measuring the spectrum has by far dominated WT studies because this is
the most basic and robust theoretical object and because this quantity
is easier to observe experimentally that more subtle statistical
quantities.  For the surface gravity waves, the WT prediction about
the energy cascade spectrum have been confirmed in several recent
numerical studies \cite{onorato,naoto,DKZ}. Here, we also start by
presenting the spectrum.  Figure \ref{EnergySpectrum} shows the
spectrum at $t=300 T_{p}$ at $t=2000T_{p}$, where $T_{p}$ is period of
the slowest mode ($k=10$).  The obtained spectrum is in agreement with
the $k^{-4}$ shape predicted by WT theory in the inertial range, and
this serves as a validation of our code.  Our subsequent statistical
measurements will be made at the time where this steady state spectrum
has already got established.
%

\subsection{Wave-amplitude probability density function and its moments}

Now we consider the PDF of amplitudes for which predictions where made
recently within the WT approach.  To measure the PDF of amplitudes
$\left| a_{\mathbf{k}}(t)\right| ^{2}$, we set two radial regions in
$\mathbf{k}$-space $k_{15}=[13,17],$ and $k_{35}=[33,37]$.  These
regions were inside the inertial range and had well mixed phases and
amplitudes since the experiment was done after performing 2000
rotations of the peak mode. We looked at the time-span of
approximately 855 rotations of modes $k=15$ or 1230 rotations of modes
$k=35$ and collected amplitudes of all modes from these three
regions. The number of amplitudes collected was over 1.1 million in
region $k_{15}$ and over 2.1 million in region $k_{35}$.

%
\begin{figure}
\begin{center}
\includegraphics[width=10cm]{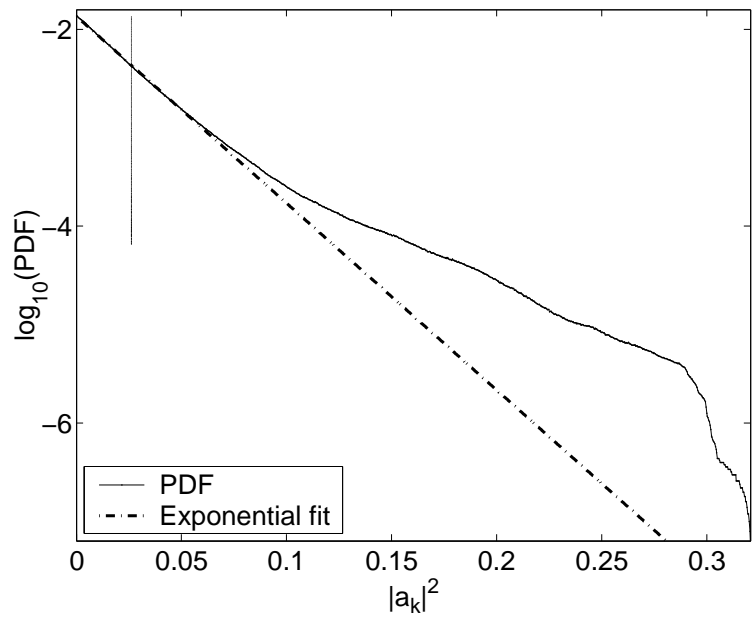}\\
\caption[6pts]{Probability density function for the amplitude
$|a_k|^2$ with $ k \in [13,17]$.  The linear fit is shown based on the
slope of the low-amplitude part (the Gaussian core).  The vertical
straight line marks the mean value (spectrum) $n_k = \langle |a_k|^2
\rangle$.}
\label{PDF15}
\end{center}
\end{figure}
%
\begin{figure}
\begin{center}
\includegraphics[width=10cm]{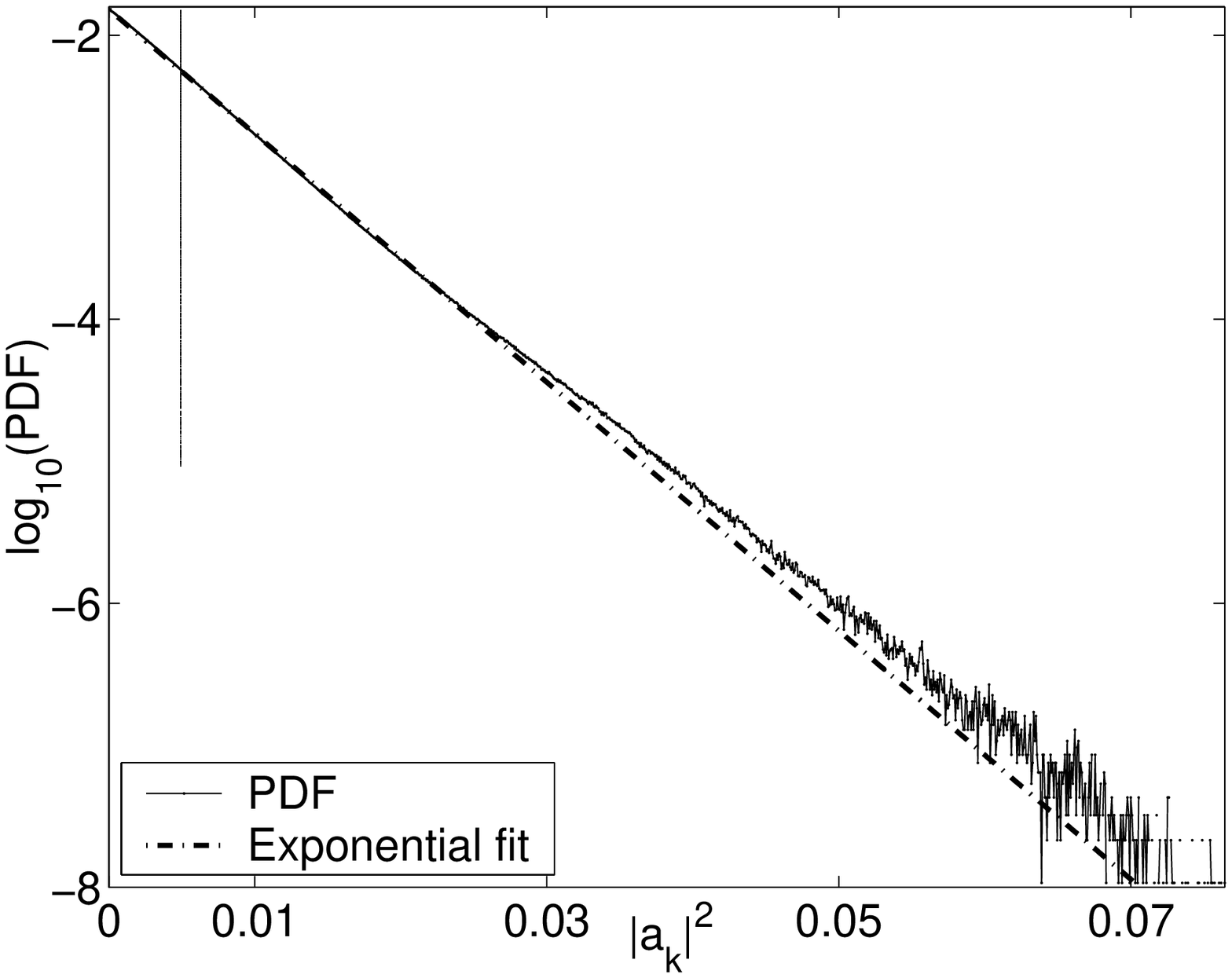}\\
\caption[6pts]{Probability density function for
the amplitude $|a_k|^2$ with $ k \in [33,37]$.
Same notations as in Figure \ref{PDF15}.}
\label{PDF35}
\end{center}
\end{figure}
Figure \ref{PDF15} shows a log-plot of the PDF the amplitude $A_k^2$
in $k_{15}$ and an exponential fit of its low-amplitude part.  One can
see intermittency, i.e., an anomalously large probability of strong
waves. We can also see that this discrepancy from Gaussianity happens
in the tail, i.e.  well below the mean amplitude value $s=n_k$.  While
the PDF tail in not long enough for drawing decisive conclusions about
realization of the theoretically predicted $1/s$ scaling, it certainly
gives a conclusive evidence that the probabilities of large amplitudes
are orders of magnitude higher than in Gaussian turbulence.  Figure
\ref{PDF35} shows the PDF of $A_k^2$ in $k_{35}$.  We can see some
non-Gaussianity in $k_{35}$ as well, although much less than in
$k_{15}$.  Similar conclusion that the gravity wave turbulence is more
intermittent at low rather than high wavenumbers was reached on the
basis of numerical simulations in \cite{naoto}.

\begin{figure}
\begin{center}
\includegraphics[width=10cm]{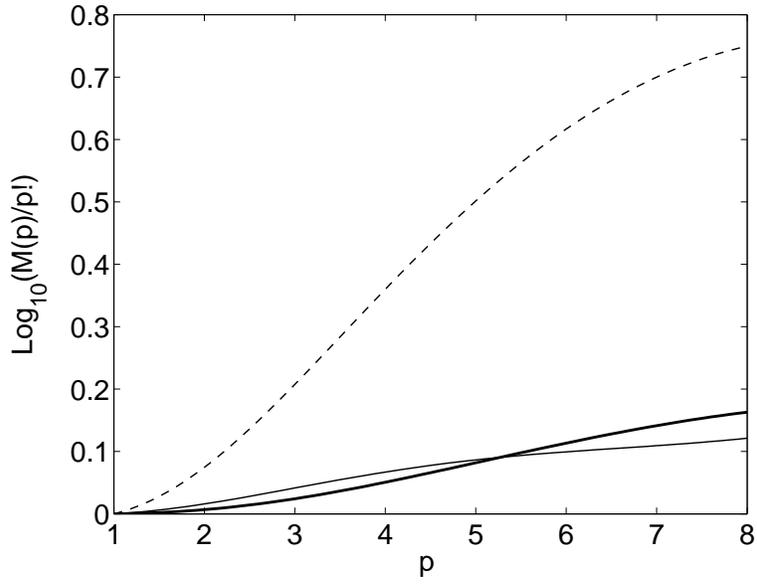}\\
\caption[6pts]{Ratio of the moments $M^{(p) } = \langle |a|^{2p} \rangle$ 
to their values in Gaussian turbulence, $n \, p!$, 
for 
$ k \in [13,17]$ (dashed line),
$ k \in [23,27]$ (thin solid line) and 
$ k \in [33,37]$ (thick solid line).
}
\label{moments}
\end{center}
\end{figure}

Deviations from Gaussianity can be also seen in figure \ref{moments}
which shows ratio of the moments $M^{(p) } = \langle |a|^{2p} \rangle$
to their values in Gaussian turbulence, $n \, p!$.  Again, we can see
that such deviations are greater at the small-$k$ part of the inertial
range.

\subsection{Frequency properties.}

\begin{figure}
\begin{center}
\includegraphics[width=10cm]{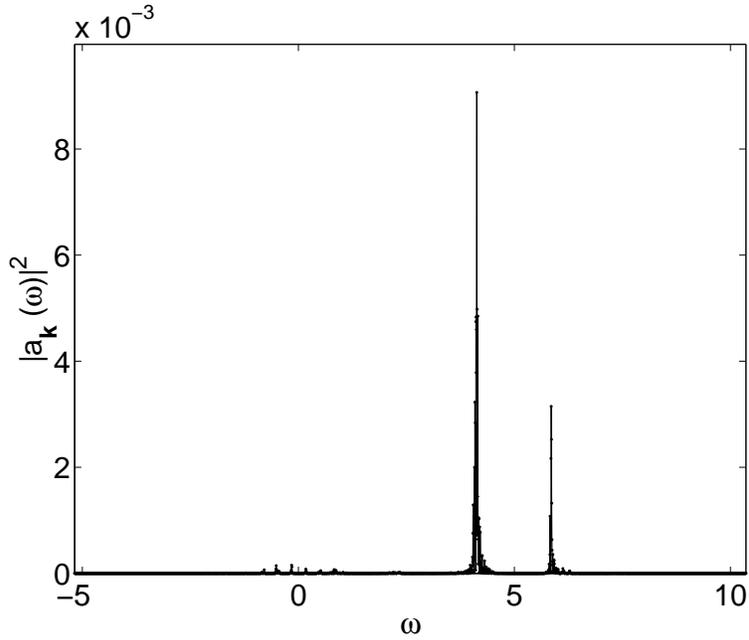}\\
\caption[6pts]{Frequency distribution of waveaction at a
fixed wavenumber ${\bf k} = (17,0)$.}
\label{freq17}
\end{center}
\end{figure}

We examine the frequency properties of waves by performing the
time-Fourier transform at each fixed wavenumber.  A typical plot, for
${\bf k} = (17, 0)$, is shown in Figures \ref{freq17}.  Our first
observation is that we always see two peaks - the bigger one at the
linear frequency and a smaller peak at a shifted frequency.  We
interpret the second peak as a nonlinear effect since there is no
frequency shift in the linear system.

%

Also, it appears that the measured ratio of squares of the peak
frequencies is approximately equal to 2 for all wavenumbers (within
10\% accuracy).  This can be explained by the nonlinear term in the
canonical transformation (\ref{nv}), e.g.  $0(\epsilon)$-term which is
quadratic with respect to the wave amplitude.  In particular, the mode
${\bf k}/2$ makes contribution to this term which oscillates at
frequency $2 \omega({\bf k}/2) = \sqrt{2 k}$ which appears to coincide
with the second peak's frequency. Thus we see that contribution of
${\bf k}/2$ dominates in the nonlinear term of the canonical
transformation.

The two-frequency character at each wavenumber has an interesting
relation to the amplitude and phase dynamics as will be seen in the
next section.

\subsection{Amplitude and phase evolution.}

%
\begin{figure}
\begin{center}
\includegraphics[width=10cm]{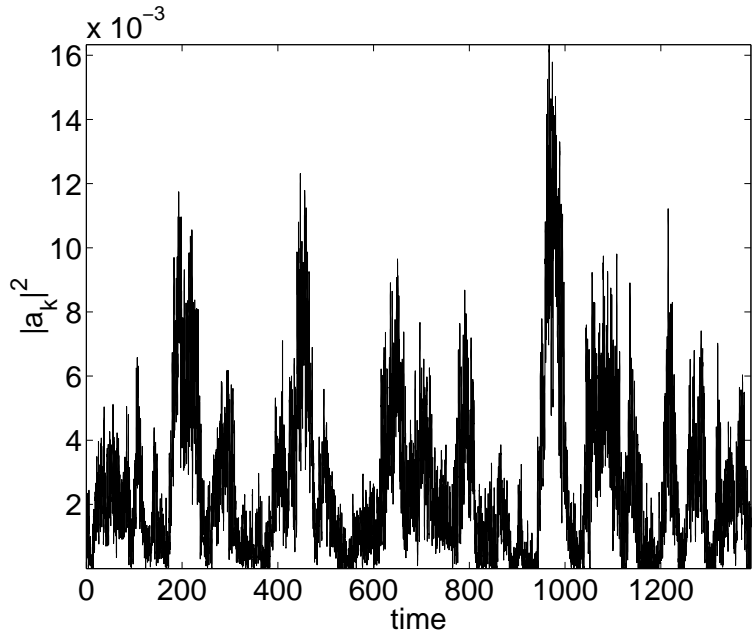}\\
\caption[6pts]{Graph of  $A_k^2(t)$
at wavenumber ${\bf k} = (25,0)$.}
\label{amp}
\end{center}
\end{figure}
%
%
%
\begin{figure}
\begin{center}
\includegraphics[width=10cm]{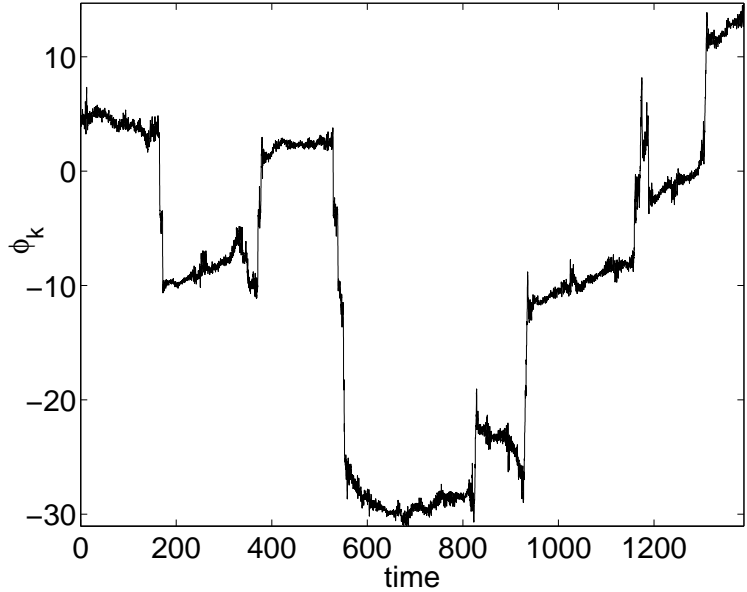}\\
\caption[6pts]
{Phase evolution, $\phi_k(t)$,
at wavenumber ${\bf k} = (25,0)$.}
\label{phase}
\end{center}
\end{figure}
%
\begin{figure}
\begin{center}
\includegraphics[width=10cm]{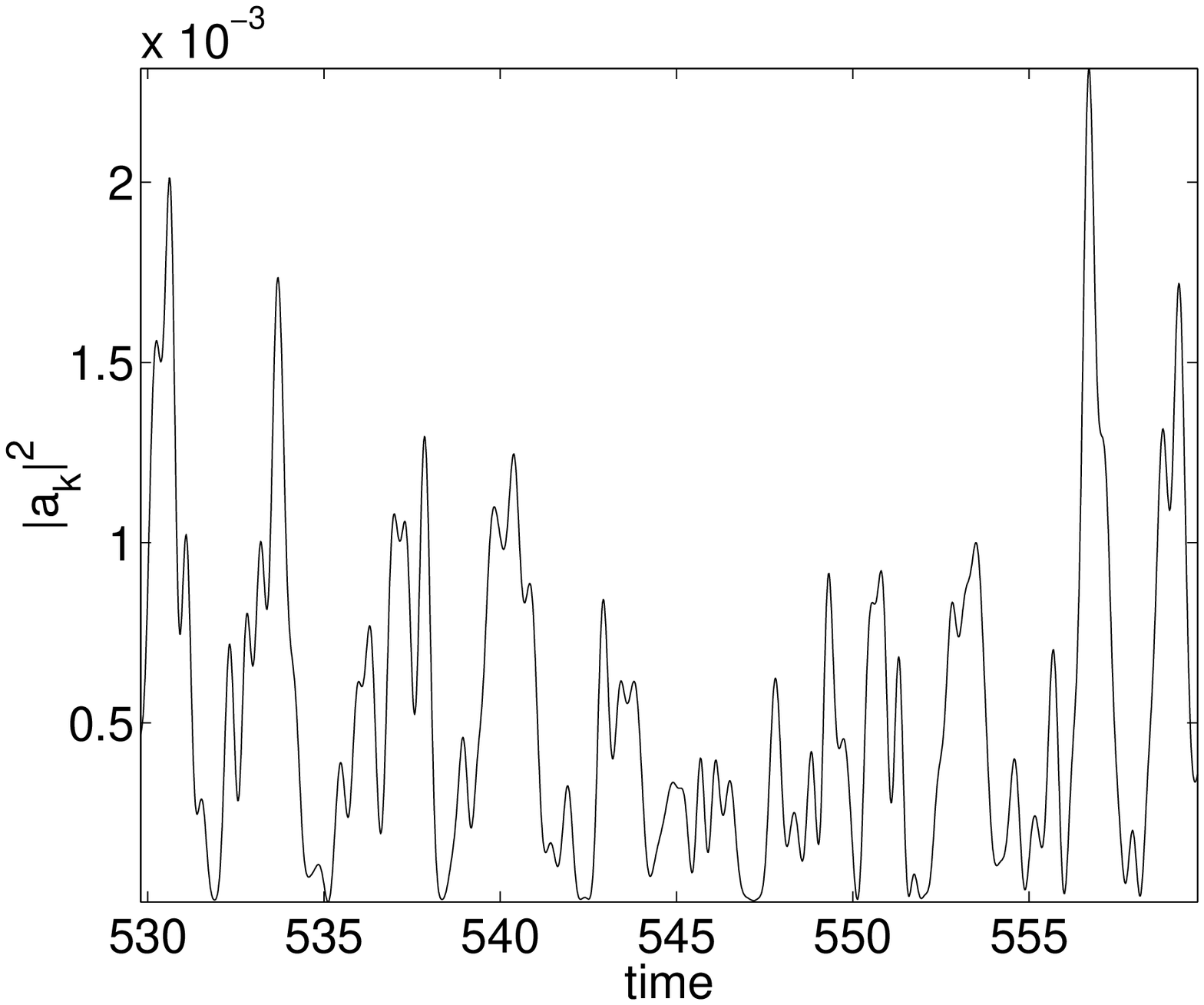}\\
\caption[6pts]{Detailed graph of  $A_k^2(t)$
at wavenumber ${\bf k} = (25,0)$ corresponding to a time
interval characterized by low amplitudes and, therefore,
phase runs. The linear wave period for this mode is $2 \pi/\omega_k
\approx 1.256$
and the characteristic nonlinear time is of the same
order of magnitude.}
\label{amp_zoom}
\end{center}
\end{figure}

Figures \ref{amp} and \ref{phase} show time evolution of the amplitude
$A_k$ and the (interaction representation) phase $\phi_k$ respectively
for ${\bf k} = (17, 0)$.  The phase evolution seen in figure
\ref{phase} consists of the time intervals when it oscillates
quasi-periodically (with amplitude less than $2 \pi$) which are
inter-leaved by sudden ``phase runs'', - fast monotonic phase changes
by values which can significantly exceed $2 \pi$.  By juxtaposing
figures \ref{amp} and \ref{phase}, one can see that when the phase
runs happen then the amplitude is close to zero.  We zoom in at the
amplitude graph in a time interval characterized by low amplitudes
(and therefore phase runs) in figure \ref{amp_zoom}.  One can see
large-amplitude quasi-periodic oscillations on $A_k$, - it changes in
value several-fold over a time comparable to the linear wave
period. This indicates that such phase run intervals mark the places
where the WT assumption of weak nonlinearity breaks down.

This behavior, together with the two-peak character of the
time-Fourier spectrum, suggest that at each wavenumber ${\bf k}$ there
are two modes:
$$
a_k = c_1  + c_2 e^{-i \omega^* t},
$$
where $\omega^* > \omega_k$ and the complex amplitudes are
$ c_1$ and $ c_2$ varying in time slower than the linear oscillations.
Most of the time $ |c_1| > |c_2|$ and, therefore, the
phase $\phi_k$ oscillates periodically about the some mean value
(equal to the phase of $ c_1$). However, sometimes $c_2$ becomes
greater than $ c_1$ and then the path of $a_k$ 
will encircle zero in the complex plane, so that 
$\phi_k$ starts gaining $2 \pi$ for each rotation.
These are the phase runs, and in order to trigger these runs
the complex-plane path of $a_k$ must encircle zero,
 which explains the observed small values of this
quantity during the phase runs. This effect is
related to the general phenomenon of phase singularities
of complex fields at points of zero amplitude (e.g. \cite{berry}).
However, our phase runs continue
all the time until  $c_2$ becomes
greater than $ c_1$ and during this time interval (excluding
its ends) the amplitude
is non-zero and the phase is perfectly regular.

Note that such observed behavior of the phase is very dependent on the
wave variables: it occurs in natural variables such as the surface
height and velocity but is absent for the waveaction variable after
the canonical transformation leading to the Zakharov equation. Thus,
the phase runs can say nothing about the dynamics or statistics of the
waveaction amplitude used in WT, but they mark short time intervals
when nonlinearity fails to be weak for a particular $k$-mode, and this
will be used as a diagnostic tool in our simulations.

\begin{figure}
\begin{center}
\includegraphics[width=10cm]{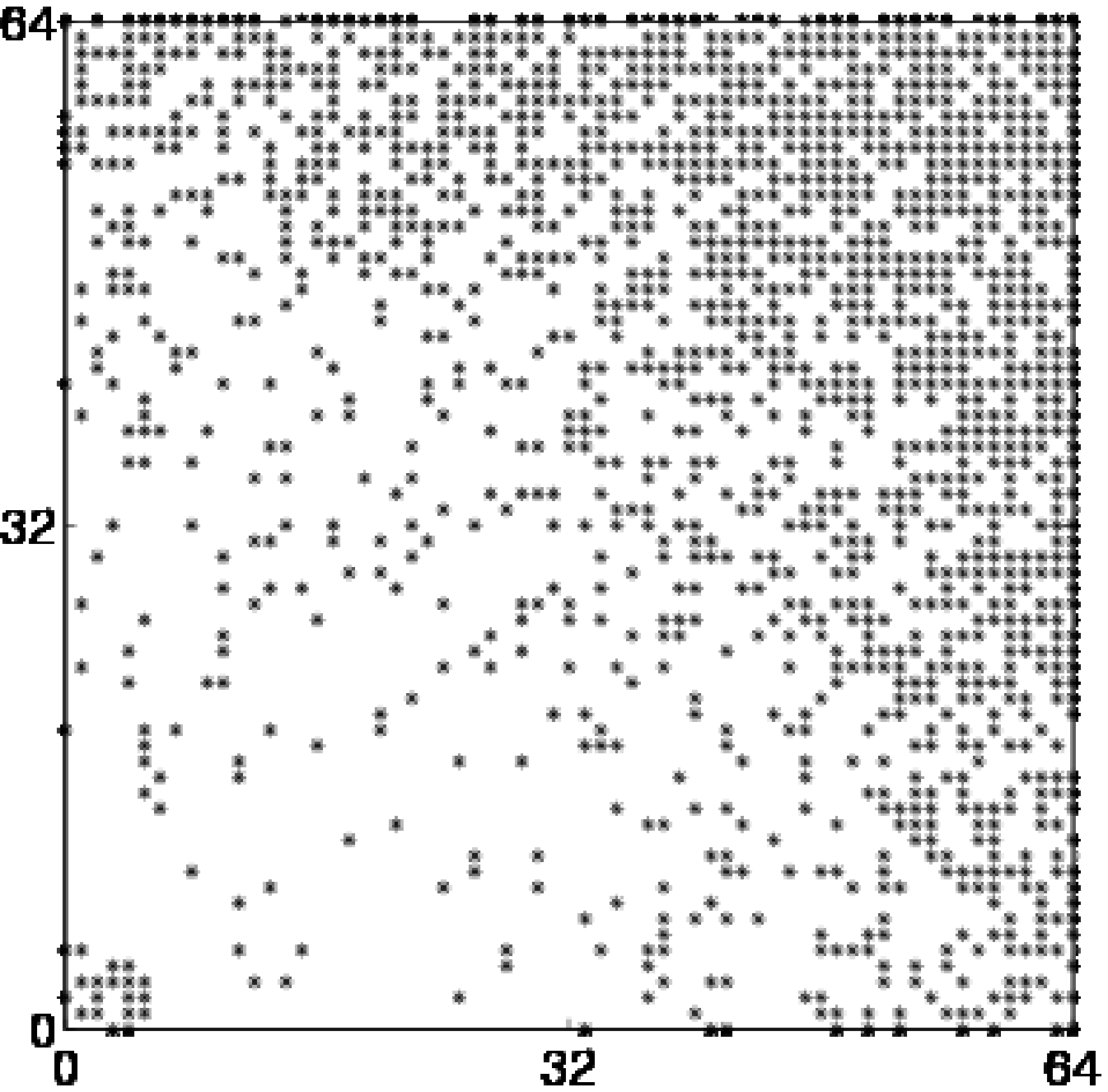}\\
\caption[6pts]
{Phase runs detected in the 2D wavenumber grid
over the time period.}
\label{phase_runs}
\end{center}
\end{figure}

\begin{figure}
\begin{center}
\includegraphics[width=10cm]{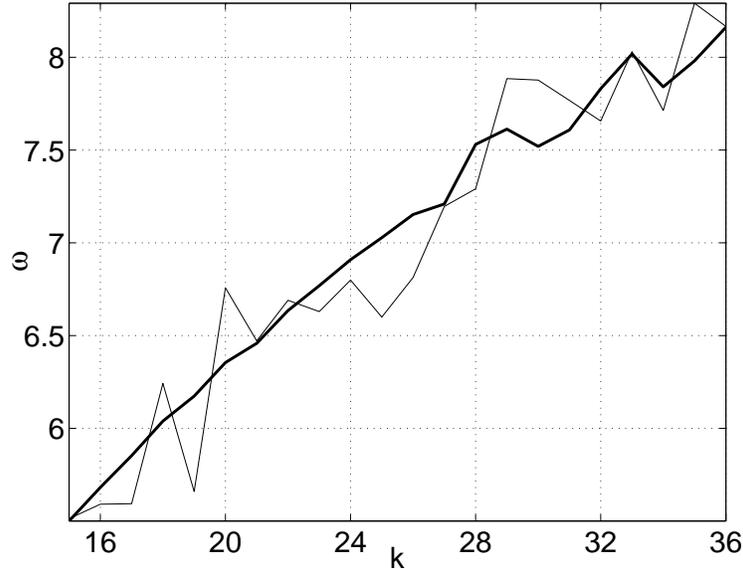}\\
\caption[6pts]
{ Mean
rate of the phase change  during the upward phase runs (thick curve)
and the second peak's frequency (thin curve) as functions of wavenumber.}
\label{upperuns}
\end{center}
\end{figure}

In figure \ref{upperuns} we show comparison of the mean
rate of the phase change  during the upward phase runs
and the second peak's frequency at different wavenumbers.
A significant coincidence of these two curves supports
the proposed above two-mode explanation.  

\begin{figure}
\begin{center}
\includegraphics[width=10cm]{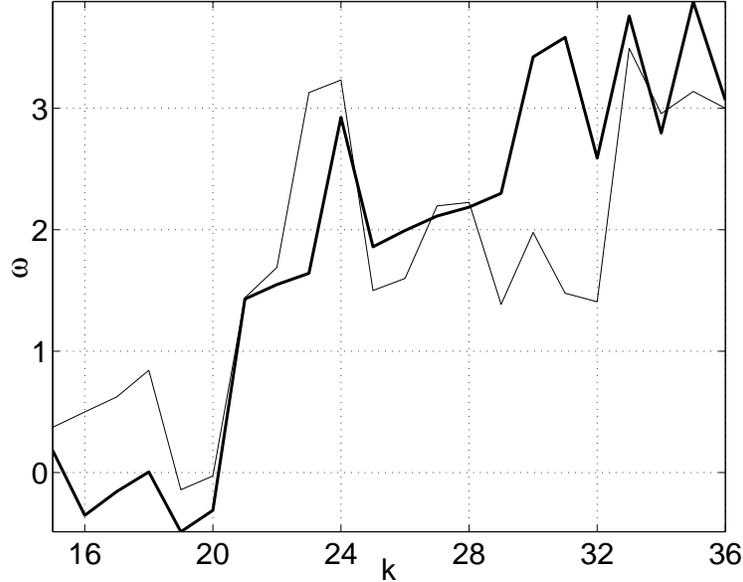}\\
\caption[6pts] { Mean rate of the phase change during the downward
phase runs (thick curve) and the frequency of the highest sub-linear
peak (thin curve) as functions of wavenumber.}
\label{loweruns}
\end{center}
\end{figure}

\begin{figure}
\begin{center}
\includegraphics[width=10cm]{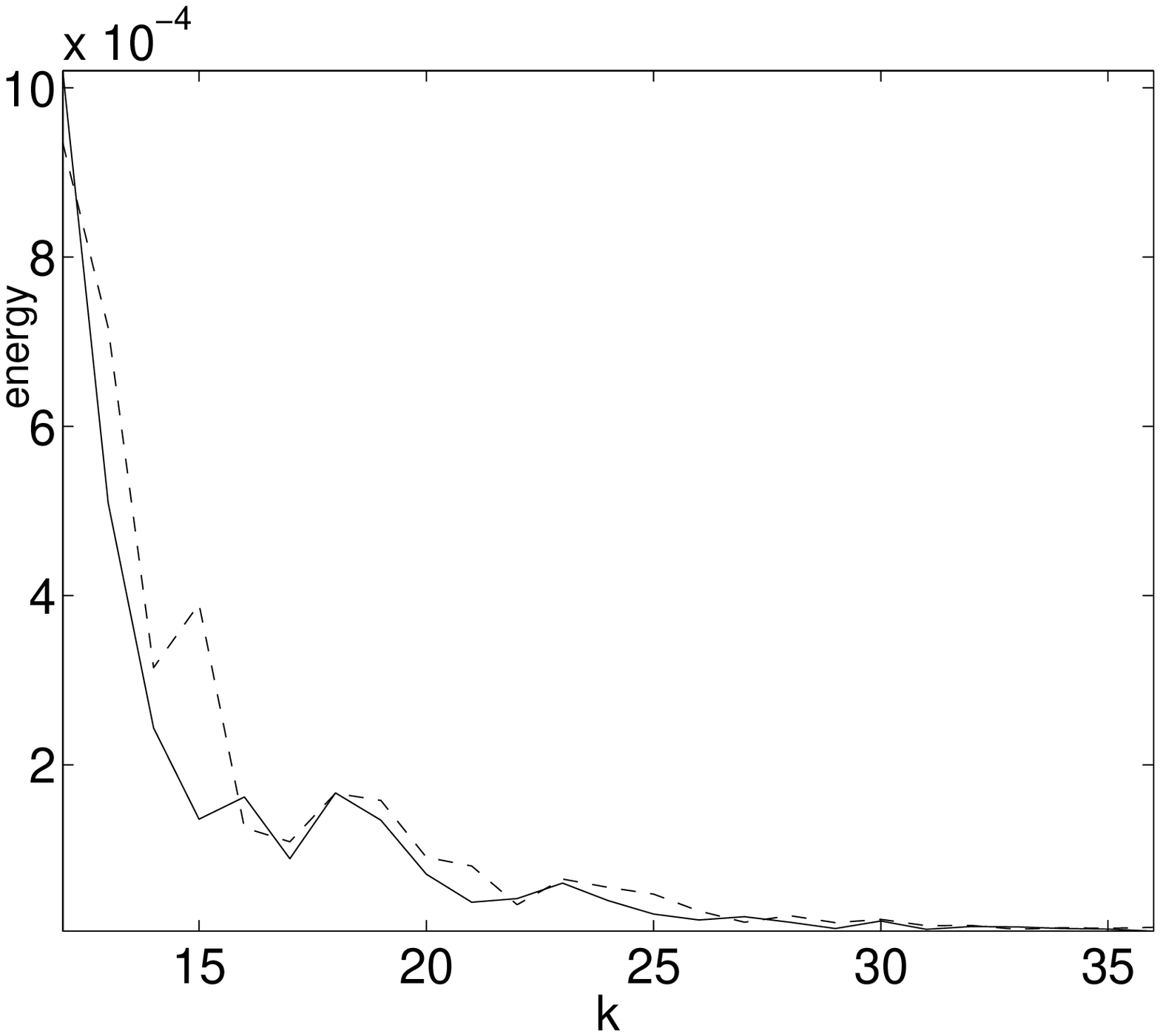}\\
\caption[6pts]
{Total energy of modes with frequencies below the linear
frequency  (solid line) and the modes with frequencies
above the linear frequency (dashed line).}
\label{below_above}
\end{center}
\end{figure}

A word of caution is due about the simple two-mode explanation of the
phase runs. Indeed, according to this picture the phase should always
run to higher values whereas in figure \ref{phase} we see both upward
and downward runs.  A possible explanation of this is that phase runs
may be triggered not only by sharp peaks but also by broadband
distributions with frequencies less than the linear one.  Such
broadband distributions could be made of modes which are strong (and
therefore can produce phase runs) but whose duration in time is short
and sporadic and at different frequencies (hence a broad
spectrum). This picture is supported by the fact that the amount of
total wave energy in the frequency range below the linear frequency is
similar (and for low wavenumbers even greater) than the amount of
energy in the range above the linear frequencies, see figure
\ref{below_above}.  Evidence of relation of the sub-linear modes and
the downward phase runs is shown in figure \ref{loweruns} where the
mean rate of the phase change during the downward phase runs and the
frequency of the highest sub-linear peak.  Again, we can see agreement
of these two curves although with a greater level of fluctuations due
to the fact the sub-linear modes are very spread over different
frequencies.

\subsection{Nonlinearly active modes and cascade ``avalanches''}

Component with the shifted frequency $\omega^*$ is clearly a nonlinear
effect (there is no frequency shift in linear dynamics).  Thus, the
relative strength of $c_2$ and $ c_1$ can be used as a measure of
nonlinearity. Particularly, the phase runs mark the events when
nonlinearity becomes strong.  Figure \ref{phase_runs} shows locations
of the phase runs in the 2D wavenumber space which happened at
$t=500$. Note that at that time ZF steady spectrum has already
formed. In the energy cascade range, we see that the phase run density
is increasing toward high $k$'s, which is in agreement with the WT
prediction that the nonlinearity grows as one cascades down-scale
\cite{rough,biven}.  Curiously, we also observe high density of the
phase runs within th circle $k<6$, which is, perhaps, manifestation of
a waveaction accumulation via an inverse cascade process. However,
this range is too small for any meaningful conclusions to be made
about the inverse cascade properties.
 
\begin{figure}
\begin{center}
\includegraphics[width=10cm]{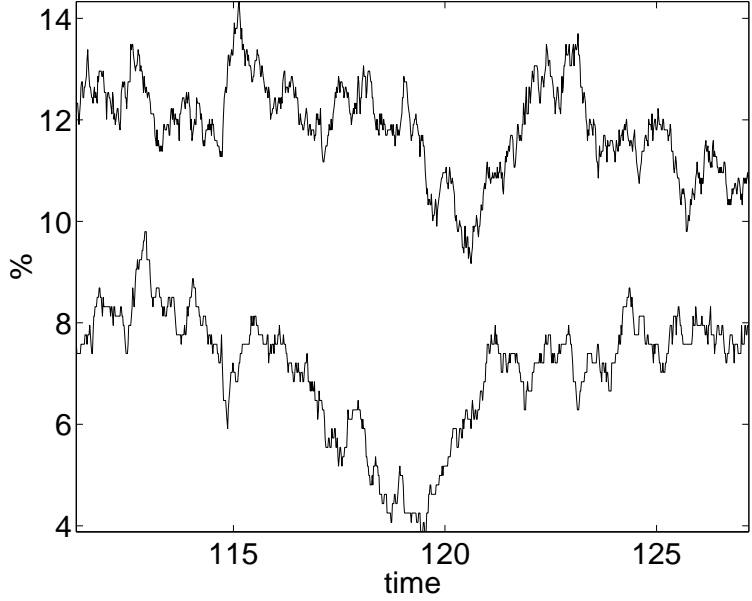}\\
\caption[6pts]
{Percentage of modes experiencing phase runs 
in the range $13 < |k| < 29$
(lower curve) and in the range $30 < |k| < 45$ (upper curve).}
\label{sandpile}
\end{center}
\end{figure}

The energy cascade from the forcing region toward the high wavenumber
region proceeds in a non-uniform in time fashion somewhat resembling
sporadic sandpile avalanches. This arises due to the $k$-grid
discreteness effects which tend to block the resonant wave interaction
when the wave intensities are small.  This situation resembles
``frozen turbulence'' of \cite{frozen}. Thus, the wave energy does not
cascade to high wavenumbers and it tends to accumulate near the
forcing scales until the wave intensity is strong enough to restore
the resonant interaction via the nonlinear resonance broadening. At
this moment the energy cascade toward high wavenumbers sets in, and
this leads to depletion of energy at the forcing scale, - ``sandpile
tips over''.  In turn, depletion of energy at the forcing scale leads
to blocking of the energy cascade, and the process continues in a
repetitive manner. As a result system oscillates between the state of
``frozen turbulence'' and the state of ``avalanche cascade''. This
behavior is illustrated in figure \ref{sandpile} which shows
percentage of modes experiencing phase runs in two different
wavenumber ranges $13 < k < 29$ and $30 < k < 45$. One can see that
the shapes of these two curves bear a great degree of similarity up to
a certain time delay and a vertical shift in the second curve with
respect to the first one.  The vertical shift reflects the fact that
the energy cascade gets stronger as it proceeds to large wavenumbers.
The time delay, on the other hand indicates the direction and the
character of the sporadic energy cascade.  It shows that a higher
(lower) nonlinear activity at low $k$'s after a finite delay causes a
higher (lower) activity at higher $k$'s, which could be compared with
propagation of an avalanche (quenching) down a sandpile.

\subsection{Correlations of phases, phase factors and amplitudes.}

\begin{figure}
\begin{center}
\includegraphics[width=10cm]{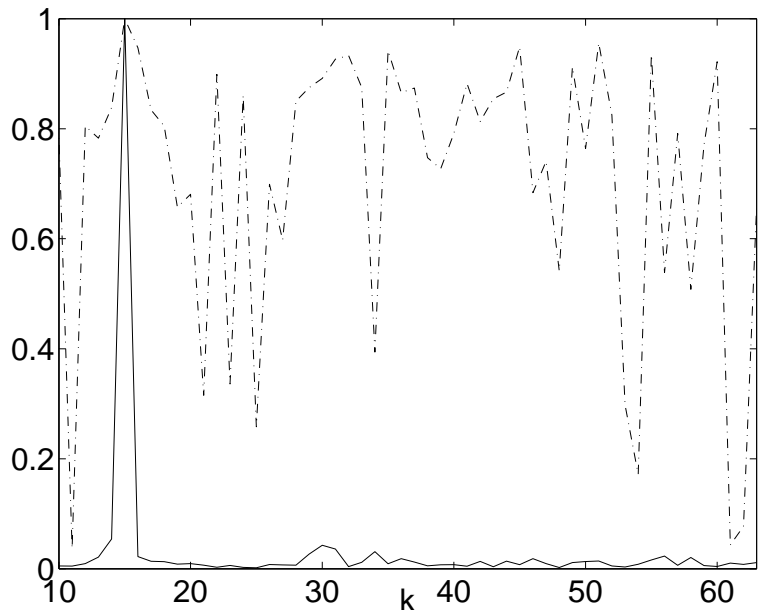}\\
\caption[6pts]
{Two-point auto-correlations for the phases ${\cal C}_{\phi,\phi}({\bf k_1},
{\bf k})$
 (dashed line) and
for the phase factors 
${\cal C}_{\psi,\psi}({\bf k_1},        
{\bf k})$
(solid line) with one point
fixed at ${\bf k} = (15,0)$.}
\label{psi_psi_corr}
\end{center}
\end{figure}

\begin{figure}
\begin{center}
\includegraphics[width=10cm]{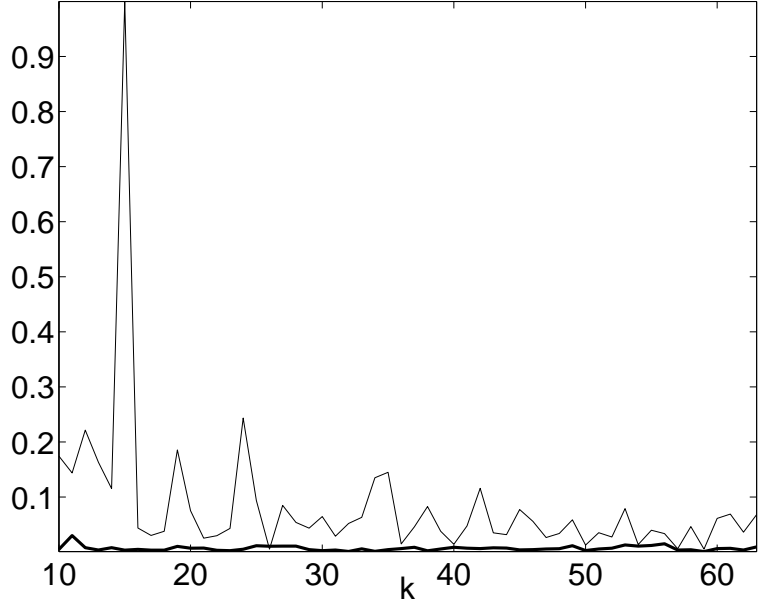}\\
\caption[6pts]
{Two-point auto-correlation of amplitudes ${\cal C}_{A,A}({\bf k_1}, 
{\bf k})$ (thin curve)
and two-point correlation between  the amplitudes and
phase factors  ${\cal C}_{A,\psi}({\bf k_1},            
{\bf k})$  (thick curve) with one point
fixed at ${\bf k} = (15,0)$.}
\label{A_psi_corr}
\end{center}
\end{figure}

WT closure relies on the RPA properties of the wave fields, i.e.  that
the amplitudes $A_k$ and the phase factors $\psi_k$ are statistically
independent variables. On the other hand, WT calculation for the
phases $\phi_k$ shows that these quantities get correlated.  In order
to check these properties and predictions numerically, let us
introduce a function that measures the degree of statistical
dependence (or independence) of some Fourier-space variables $ X({\bf
k}_1)$ and $ Y({\bf k}_2)$,
\begin{eqnarray}
{\cal C}_{X,Y} ({\bf k}_1, {\bf k}_2) &=& 
\frac{ \langle X({\bf k}_1) Y({\bf k}_2) \rangle - \langle X({\bf k}_1)
 \rangle \langle Y({\bf k}_2) \rangle }{
\sqrt{ \langle  X^2({\bf k}_1) \rangle -  \langle X({\bf k}_1) \rangle^2   } 
\sqrt{ \langle  Y^2({\bf k}_2) \rangle -  \langle Y({\bf k}_2) \rangle^2   } }.
 \end{eqnarray}
For example, we can examine to what degree amplitudes $A$ and independent
of the phase factors $\psi$ by looking at the function
${\cal C}_{A,\psi} ({\bf k}_1, {\bf k}_2)$ for different values of ${\bf k}_1$
and ${\bf k}_2$. Independence of  the amplitudes at different wavenumbers
can be examined by the auto-correlation function
${\cal C}_{A,A} ({\bf k}_1, {\bf k}_2)$, and similar for the phase factors
and the phases.
We restrict ourselves with choosing ${\bf k}_1 =(15,0)$ and
${\bf k}_2 =(k,0)$ with $k \in (10,64)$. 
Figure \ref{psi_psi_corr} shows the values of correlators
${\cal C}_{\phi,\phi} ({\bf k}_1, {\bf k}_2)$ and ${\cal C}_{\psi,\psi} ({\bf k}_1, {\bf k}_2)$
 as functions of $k$.
In agreement with WT predictions, 
auto-correlations of   $\psi_k$'s are very small
 whereas
the ones of  $\phi_k$'s are significant (except, of course,
for $k=15$ where by definition these correlators are equal to one).
Correlators
${\cal C}_{A,A} ({\bf k}_1, {\bf k}_2)$ and ${\cal C}_{A,\psi} ({\bf k}_1, {\bf k}_2)$
are shown in figure \ref{A_psi_corr}. Again, we see a good agreement 
with the WT prediction: these correlations are  very small (except, again,
${\cal C}_{A,A}(15,15)=1$).

%

\section{Discussions}

In this paper, we used direct numerical simulations of the free water
surface in order to examine the statistical properties of the
water-wave field beyond the energy spectrum.  Our first aim was to
check recent predictions of the WT theory about the PDF and
intermittency, about the character of correlations of the wave
amplitudes and phases.  We particularly focused on the question how
the effects of discreteness and finite nonlinearity change statistics
with respect to the WT closure developed for weak nonlinearities and
for a continuous wavenumber space.

Firstly, following \cite{DKZ,onorato,naoto} we see formation of a
quasi-steady spectrum consistent with
the Zakharov-Filonenko spectrum predicted by WT. Secondly, we measured
PDF for the wave amplitudes and observe an anomalously large, with
respect to Gaussian fields, probability of strong waves. This result
is in agreement with recent theoretical predictions of
\cite{clnp,cln}.  Thirdly, we measure correlations for the amplitudes,
phases and phase factors and we observe agreement with predictions of
\cite{clnp,cln}. Namely, the amplitude and the phase correlations
behave as statistically independent variables, whereas the phases
develop strong auto-correlations over the nonlinear time.  Note that
these properties are fundamental for the WT closure to work, so in a
way we provide a numerical validation for the WT approach.

We also find that at each $k$ there are two sharp frequency peaks: a
dominant one at the linear frequency and a weaker one with a frequency
shift arising due to the $k/2$-mode.  Somewhat related to this
two-peak frequency structure is the observed time behavior of the
phase.  We observe calm periods during which the phase oscillates
within $2 \pi$-wide margins intermittent with sudden phase runs during
which it experiences a monotonic change significantly greater than $2
\pi$.

Finally, we observe that the energy cascade is ``bursty'' in time and
is somewhat similar to sporadic sandpile avalanches. We give a
plausible explanation of this behavior as an interplay of effects of
discreteness and nonlinearity. Because in between of the avalanche
discharges the resonances are absent then, at least qualitatively, one
can refer to the KAM theory and say that the evolution should remain
close to the corresponding integrable case, - the linear system in our
case.  This picture is supported by a simple analysis of
quasi-resonances given in this paper which indicates that there exists
a single threshold value of turbulence intensity at the forcing scale
separating the no-cascade and unlimited (in $k$) cascade regimes.

A further numerical study of the avalanche effect is desirable,
particularly using a different wavenumber grid and using a more direct
method of measuring the turbulent flux and its correlations for
different inertial interval points.

\section{Acknowledgments}
We thank Miguel Onorato and Victor Shrira for helping us to understand
the nature of the second frequency peak, and Miles Reid for teaching
us some Number Theory basics.

\end{document}